\documentclass[superscriptaddress,11pt]{revtex4}

\usepackage{amssymb,graphicx,amsmath,array,verbatim,ascmac,amsthm}

\newcommand{\beq}{\begin{eqnarray}}
\newcommand{\eeq}{\end{eqnarray}}
\newcommand{\p}{\partial}

\newcommand{\vs}[1]{\vspace{#1 mm}}
\newcommand{\hs}[1]{\hspace{#1 mm}}
\newcommand{\bpm}{\begin{pmatrix}}
\newcommand{\epm}{\end{pmatrix}}
\newcommand{\Z}{\mathbb{Z}}
\newcommand{\R}{\mathbb{R}}
\newcommand{\C}{\mathbb{C}}
\newcommand{\tr}{{\rm Tr}}
\newcommand{\D}{\mathcal D}

\newcommand{\ba}{\left(\begin{array}}
\newcommand{\ea}{\end{array} \right)}

\makeatletter
\@addtoreset{equation}{section}

\makeatother

\begin{document}

\title{
Non-perturbative Contributions from Complexified Solutions
in ${\mathbb C}P^{N-1}$ Models
}

\author{Toshiaki Fujimori}
\email{toshiaki.fujimori018(at)gmail.com}
\address{Department of Physics, 
Keio University, 4-1-1 Hiyoshi, Yokohama, Kanagawa 223-8521, Japan
}
\address{Research and 
Education Center for Natural Sciences, 
Keio University, 4-1-1 Hiyoshi, Yokohama, Kanagawa 223-8521, Japan
}

\author{Syo Kamata}
\email{skamata(at)rikkyo.ac.jp}
\address{Research and 
Education Center for Natural Sciences, 
Keio University, 4-1-1 Hiyoshi, Yokohama, Kanagawa 223-8521, Japan
}

\author{\\ Tatsuhiro Misumi}
\email{misumi(at)phys.akita-u.ac.jp}
\address{Department of Mathematical Science, Akita 
University, 1-1 Tegata Gakuen-machi,
Akita 010-8502, Japan
}
\address{Research and 
Education Center for Natural Sciences, 
Keio University, 4-1-1 Hiyoshi, Yokohama, Kanagawa 223-8521, Japan
}

\author{Muneto Nitta}
\email{nitta(at)phys-h.keio.ac.jp}
\address{Department of Physics, 
Keio University, 4-1-1 Hiyoshi, Yokohama, Kanagawa 223-8521, Japan
}
\address{Research and 
Education Center for Natural Sciences, 
Keio University, 4-1-1 Hiyoshi, Yokohama, Kanagawa 223-8521, Japan
}

\author{Norisuke Sakai}
\email{norisuke.sakai(at)gmail.com}
\address{Department of Physics, 
Keio University, 4-1-1 Hiyoshi, Yokohama, Kanagawa 223-8521, Japan
}
\address{Research and 
Education Center for Natural Sciences, 
Keio University, 4-1-1 Hiyoshi, Yokohama, Kanagawa 223-8521, Japan
}

\begin{abstract}
We discuss the non-perturbative contributions 
from real and complex saddle point solutions in the ${\mathbb C}P^{1}$ 
quantum mechanics with fermionic degrees of freedom, 
using the Lefschetz thimble formalism beyond the gaussian approximation.
We find bion solutions, 
which correspond to (complexified) instanton-antiinstanton configurations 
stabilized in the presence of the fermonic degrees of freedom. 
By computing the one-loop determinants in the bion backgrounds, 
we obtain the leading order contributions from both the real 
and complex bion solutions. 
To incorporate quasi zero modes 
which become nearly massless in a weak coupling limit, 
we regard the bion solutions 
as well-separated instanton-antiinstanton configurations 
and calculate a complexified quasi moduli integral 
based on the Lefschetz thimble formalism. 
The non-perturbative contributions from the real and complex bions 
are shown to cancel out in the supersymmetric case
and give an (expected) ambiguity in the non-supersymmetric case, 
which plays a vital role in the resurgent trans-series. 
For nearly supersymmetric situation, evaluation of the 
Lefschetz thimble gives results in precise agreement with those of 
the direct evaluation of the Schr\"odinger equation.
We also perform the same analysis for the sine-Gordon quantum mechanics and 
point out some important differences 
showing that the sine-Gordon quantum mechanics 
does not correctly describe the 1d limit of the ${\mathbb C}P^{N-1}$ field theory of $\R \times S^1$.
\end{abstract}

\maketitle

\newpage

\tableofcontents

\newpage 


\section{Introduction}
\label{sec:Intro}

The application of the resurgence theory and the framework of
the complexified path integral to quantum theories 
has recently been attracting a great deal of attention \cite{Ec1, Marino:2007te, Marino:2008ya, Marino:2008vx, Pasquetti:2009jg, Drukker:2010nc, Aniceto:2011nu, Argyres:2012vv, Argyres:2012ka, Marino:2012zq, Dunne:2012ae, Dunne:2012zk}.
In the resurgence theory, all the perturbative series around nontrivial backgrounds are taken into account, and it is expected that a full semi-classical expansion in 
perturbative and non-perturbative sectors, which is called a ``resurgent" trans-series,
leads to unambiguous and self-consistent definition of quantum theories
\cite{Schiappa:2013opa, Hatsuda:2013oxa, Dunne:2013ada, Cherman:2013yfa, Basar:2013eka, Aniceto:2013fka,Santamaria:2013rua, Dunne:2014bca, Cherman:2014ofa, Misumi:2014jua, Grassi:2014cla, Sauzin, Couso-Santamaria:2014iia, Misumi:2014bsa, Aniceto:2014hoa, Misumi:2014rsa, Nitta:2014vpa, Couso-Santamaria:2015wga, Behtash:2015kna, Nitta:2015tua, Dunne:2015ywa, Aniceto:2015rua, Dorigoni:2015dha, Misumi:2015dua, Behtash:2015zha, Behtash:2015loa, Dunne:2015eaa, Gahramanov:2015yxk, Dunne:2016nmc, Dunne:2016qix, Honda:2016mvg, Misumi:2016fno,Honda:2016vmv}.

While the non-perturbative backgrounds in the resurgent trans-series 
usually mean exact solutions including instantons,
certain quantum theories require taking account of non-solution configurations \cite{Bogomolny:1980ur, ZinnJustin:1981dx, ZinnJustin:1982td, ZinnJustin:1983nr, ZinnJustin:2004ib, ZinnJustin:2004cg, Jentschura:2010zza, Jentschura:2011zza} as  
``bions" composed of an instanton and an anti-instanton. 
Indeed, imaginary ambiguities arising in bion contributions
cancel out those arising in non-Borel-summable perturbative series 
in the double-well and the sine-Gordon quantum mechanics \cite{Bogomolny:1980ur, ZinnJustin:1981dx, ZinnJustin:1982td, ZinnJustin:1983nr, ZinnJustin:2004ib} .
From the viewpoint of resurgence theory in ordinary differential equations (ODE)\cite{Ec1},
the reason why the non-saddle-point configurations 
play the most relevant role in the resurgent trans-series is unclear.  

The resurgence structure in the quantum theories 
gets clear if we extend the configuration space by complexifying real variables.
(Note that complexification of integration contours is a standard technique 
to perform ordinary zero-dimensional integrals.)
In terms of the framework of the complexified path integral,
the imaginary ambiguity in the Borel resummation of perturbative series 
is one of the symptoms of ``Stokes phenomena" at ${\rm arg} \, g^{2}=0$ ($g^{2}$ is a perturbation parameter).
Here the bion configuration is expected to correspond to the 
complex saddle point, to which the deformed integration
contour is attached for ${\rm arg} \, g^{2}\not=0$.
Such complex contours decomposed from the original 
integration contour are called Lefschetz thimbles, 
along which the imaginary part of the action is unchanged \cite{Witten:2010cx}.
Thus, the contribution from each background in the resurgent 
trans-series can be seen as the integral along the Lefschetz 
thimble associated with each saddle point. 
It is notable that Stokes phenomena urge us to complexify 
the variables when ${\rm arg} \, g^{2}=0$ is the Stokes line.

Recent investigation on exact solutions of the holomorphic equations of motion 
for the complexified path integral of quantum mechanics is 
initiated in Refs.~\cite{Behtash:2015zha, Behtash:2015loa}.
In these papers, it is shown that, in the complexified path integral of double-well 
and sine-Gordon quantum mechanics with fermionic degrees of freedom, 
the bion configurations appear as the complexified exact solutions.
It indicates that the partition function of quantum theory can be defined by performing 
integration along the thimbles associated with the exact solutions.  
However, the integration along the Lefschetz thimbles has not been 
performed explicitly, even in the one-loop approximations. 

Therefore, {\it what we need to do on this topic} are summarized as follows:
(i) To calculate the one-loop contributions from the complexified solutions 
in the quantum mechanical systems with fermionic degrees of freedom.
(ii) To show that the results are consistent with the known facts in the supersymmetric cases. 
(iii) To evaluate the integral along the Lefschetz thimble 
by considering the complexified quasi moduli integral for the bion configurations.

{\it What we will show in this paper} is summarized as follows:
We obtain one-loop contributions around real and complex 
bion (instanton-antiinstanton) solutions in the ${\mathbb C}P^{1}$ and the sine-Gordon quantum mechanics 
with the parameter $\epsilon$ corresponding to fermionic degrees of freedom. 
To consider the path integral along the Lefschetz thimbles 
associated with the exact solutions, 
we regard the solutions as 
well-separated instanton-antiinstanton configurations and 
calculate the complexified quasi moduli integral along Lefschetz thimbles.
It can be viewed as a rigorous version of the Bogomolny--Zinn-Justin prescription.
For the supersymmetric parameter set, we show that the one-loop 
contributions from the real and complex solutions are cancelled out.
For nearly supersymmetric case of $\epsilon \approx 1$, 
we find that the result of Lefschetz thimble evaluation correctly reproduces 
the non-perturbative contribution to the ground state energy, 
which can be evaluated as an insertion of an operator proportional to $\epsilon-1$, 
giving the $\mathcal O(\epsilon-1)$ contribution exactly as a function of $g$.  
These non-perturbative contributions from the real and complex 
bion solutions in the ${\mathbb C}P^{N-1}$ quantum mechanics are of 
importance in the resurgent trans-series. 
We also argue that our result helps to catch the 
resurgence structure in the 2d sigma model, 
since the bions in the ${\mathbb C}P^{N-1}$ sigma model on ${\mathbb R}^{1}\times S^{1}$ 
directly reduces to those in ${\mathbb C}P^{N-1}$ quantum mechanics by dimensional reduction.

We also perform the same analysis for the sine-Gordon quantum mechanics 
and find some important differences from the $\C P^1$ case, 
which contradicts with the conjectured relevance of the sine-Gordon quantum mechanics 
to the two-dimensional ${\mathbb C}P^{N-1}$ field theory on the compactified space $\R \times S^1$. 

The organization of this paper is as follows. 
In Sec.II, we derive ${\mathbb C}P^{N-1}$ quantum mechanics 
from two-dimensional ${\mathbb C}P^{N-1}$ sigma model
and discuss (nearly-)supersymmetric cases 
by dealing with Schroedinger equation exactly.
In Sec.III, we show the real and complex bion solutions 
in complexified ${\mathbb C}P^{N-1}$ quantum mechanics and
calculate their contributions by computing the one-loop determinants in the bion background. 
In this section, we also manifest the existence of quasi zero modes around the saddle points.
In Sec.IV, we perform effective thimble integrals 
associated with the real and complex bion saddle points, 
and show that they are in good agreement 
with the exact near-supersymmetric results.
In Sec.V, we perform the same calculation for the sine-Gordon quantum mechanics, 
and show its difference from
${\mathbb C}P^{N-1}$ quantum mechanics.
Sec.VI is devoted to summary and discussion.


\section{${\mathbb C}P^{N-1}$ Quantum Mechanics from 2d field theory}
In this section, we briefly review the dimensional reduction of 
the $\C P^{N-1}$ non-linear sigma model on $\R \times S^1$ 
with a special emphasis on fractional instantons and bion configurations,
which are responsible for non-perturbative effects both in the 2d field theory and the quantum mechanics.

\subsection{2d ${\mathbb C}P^{N-1}$ non-linear sigma model }

The action of the ${\mathbb C} P^{N-1}$ model in 2d Euclidean space 
is given in terms of the inhomogeneous coordinates $\varphi^i(x)~(i=1,\cdots,N-1)$ as 
\begin{eqnarray}
S = {1\over{g^{2}_{\rm 2d}}} \int d^2 x \, G_{i \bar j} \p_\mu \varphi^i \p^\mu \bar \varphi^j \,, 
\hs{10}
G_{i \bar j} = \frac{\p^2}{\p \varphi^i \p \bar \varphi^j} \log (1 + \varphi^k\bar \varphi^k). 
\label{eq:action}
\end{eqnarray}
This model has topologically non-trivial instanton solutions characterized by the topological charge \cite{Polyakov:1975yp}
\begin{eqnarray}
Q ~=~ \frac{1}{2\pi} \int dx_1 dx_2 \, i \epsilon^{\mu\nu} G_{i \bar j} \p_\mu \varphi^i \p_\nu \bar \varphi^j \,.
\label{eq:top_charge}
\end{eqnarray}
To write down instanton configurations, 
it is convenient to use the homogeneous coordinates $h$ ($n$-component row vector) 
which specify a point in $\C P^{N-1}$ as a complex line in $\C^N$, 
i.e. the physical field corresponds to an equivalence class 
\beq
h(x) \sim V(x) h(x) \,,
\label{eq:equivalence}
\eeq
where $V(x) \in \C^\ast$ is an arbitrary nonsingular function. 
Since $h$ multiplied by any functions should be identified, 
we can choose one of the non-vanishing component to be unity. 
For example, if we fix the first component of $h$, 
the homogeneous and inhomogeneous coordinates are related as
\beq
h = (1, \varphi^1, \cdots , \varphi^{N-1}) \,.
\eeq
To discuss application of the resurgence theory to the present theory, 
it has been extremely 
useful to compactify one dimension as $ x_2 + L \sim x_2$. 
In the compactified space, one can impose the 
${\mathbb Z}_{N}$-twisted boundary condition
\beq
\varphi^k(x_1, x_2+L) = \varphi^k(x_1, x_2) e^{\frac{2\pi k i}{N}}. 
\label{eq:ZNboundCod_inhomo}
\eeq
In terms of $h(x)$, the ${\mathbb Z}_{N}$-twisted boundary condition can be rewritten as
\begin{equation}
h(x_1, x_2+L) = h(x_1,x_2) \, \Omega, \hs{10} 
\Omega={\rm diag} \left[1, e^{2\pi i/N}, \cdots, e^{2(N-1)\pi i/N} \right]. 
\label{eq:twisted_bc}
\end{equation}
In fact, it has been shown that one-loop effective potential 
for the holonomy in the compactified space favors this boundary 
condition in various situations\cite{Dunne:2012ae,Dunne:2012zk}. 
In the presence of the non-trivial holonomy, 
the continuously degenerated classical vacuum reduces to $N$ discrete points, 
each of which corresponds to $h(x)$ with only one nonzero component. 

The solution of the (anti-)BPS instanton equation $\p_{\bar z} \varphi^i = 0~(\p_{z} \varphi^i = 0)$ 
is given by an arbitrary (anti-)holomorphic function $h(z)~(h(\bar z))$ 
with respect to the complex coordinates $z=x_1+ix_2$. 
The ${\mathbb Z}_{N}$-twisted boundary conditions are satisfied 
if and only if the homogeneous coordinate $h$ takes the form
\beq
h = \left( P_1(e^{\frac{2\pi z}{L}}) \, , \, P_2(e^{\frac{2\pi z}{L}}) \, , \, \cdots \, , \, P_N(e^{\frac{2\pi z}{L}}) \right) 
{\rm diag} \left[ 1 \, , \, e^{mz} \, , \, \cdots \, , \, e^{(N-1)mz} \right], 
\eeq 
where $P_1,P_2,\cdots,P_N$ are polynomial of $\exp \left( \frac{2\pi z}{L} \right)$ 
and the parameter $m$ is given by
\beq
m ~\equiv~ \frac{2\pi}{N} \frac{1}{L}.
\eeq
The twisted boundary condition allows the fractional instanton (we denote it as ${\cal I}$) 
with topological charge $Q=1/N$ as the simplest building blocks of BPS solutions \cite{Eto:2004rz,Eto:2006mz,Eto:2006pg}.
There are $N$ types of fractional instantons, 
each of which interpolates a pair of adjacent vacua:
\begin{equation}
h_{\cal I} ~=~ \left(0 \, ,\, \cdots \, ,\, 0 \, ,\, 1 \, ,\, a \, ,\, 0 \, ,\, \cdots \, ,\, 0 \right) 
{\rm diag} \left[ 1 \, , \, e^{mz} \, , \, \cdots \, , \, e^{(N-1)mz} \right]. 
\label{eq:fractional_instanton}
\end{equation}
The fractional instanton has one complex (two real) moduli parameter $a$, 
which is related to its position $x_1= \frac{1}{m} \log |a|$ and 
intrinsic phase $\phi \equiv {\rm arg} \, a$. 

\begin{figure}[htbp]
\begin{center}
\begin{minipage}{0.45\hsize}
\centering
\includegraphics[width=75mm]{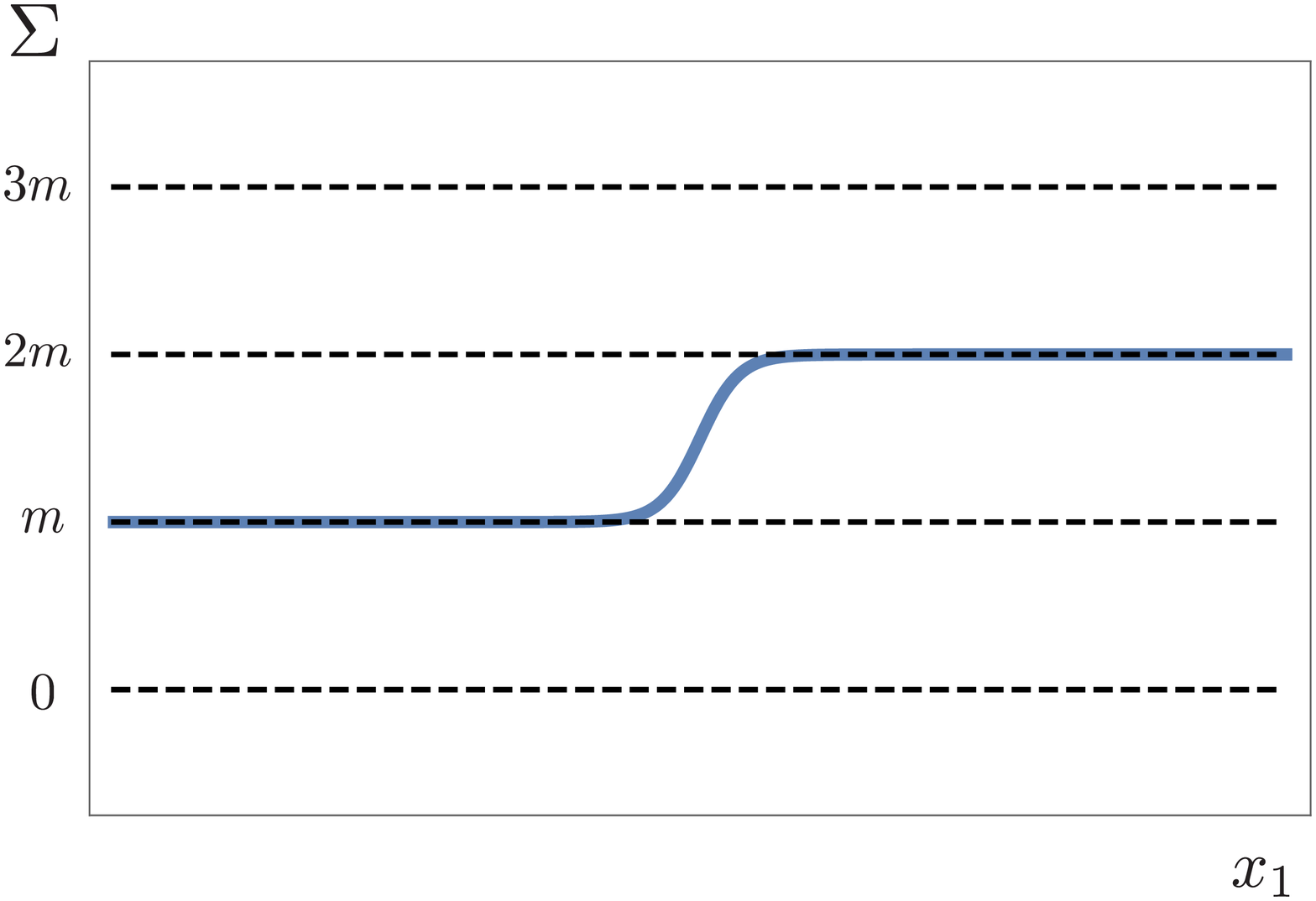} \\ \vs{-5}
(a) fractional instanton
\end{minipage}
\hs{10}
\begin{minipage}{0.45\hsize}
\centering
\includegraphics[width=75mm]{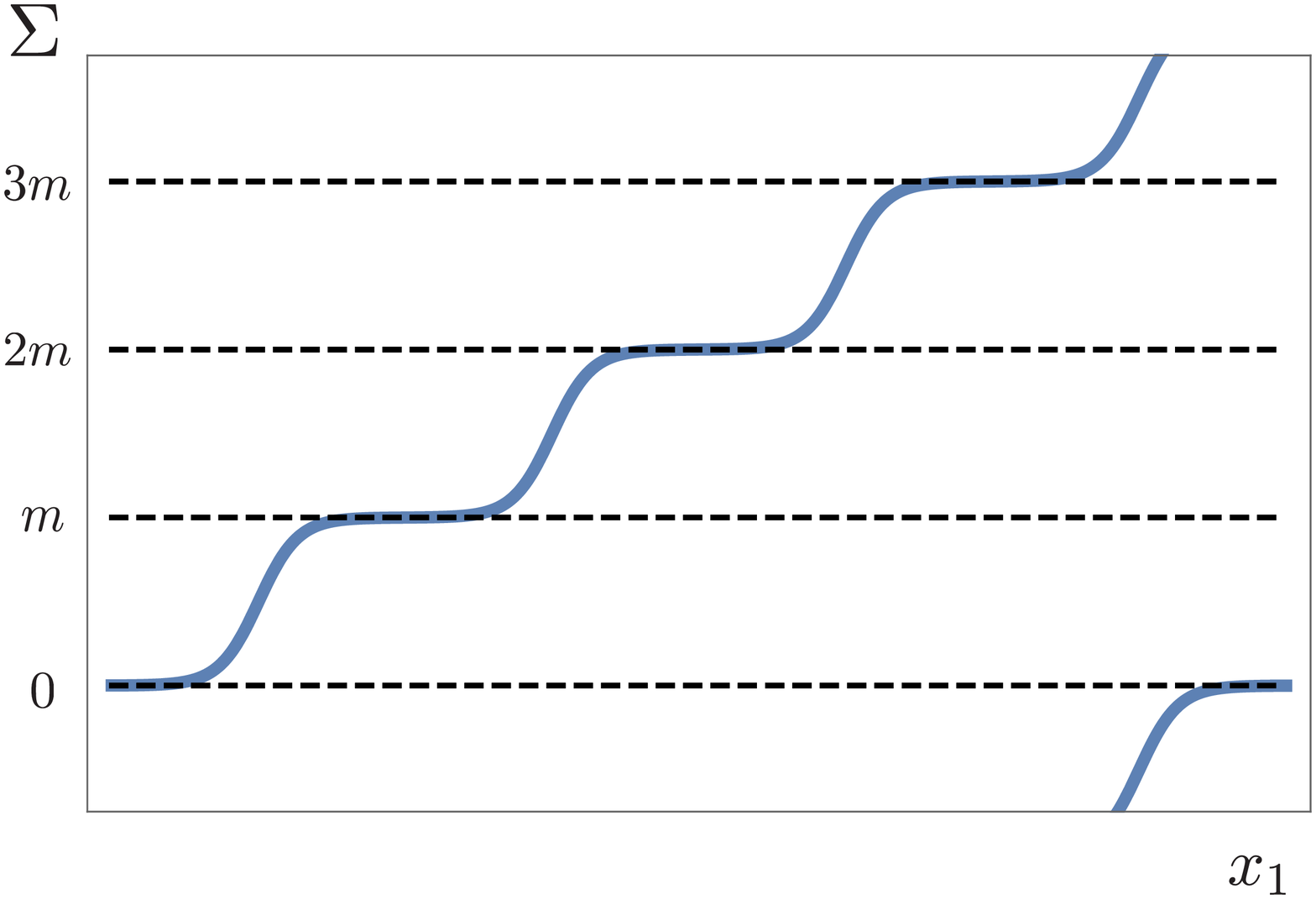} \\ \vs{-5}
(b) instanton
\end{minipage}
\caption{Kink-like profiles of (fractional) instanton configurations in ${\mathbb C}P^{3}$ model
with ${\mathbb Z}_{4}$-twisted boundary conditions. 
Each horizontal dotted line corresponds to the value of $\Sigma$ in each vacuum. 
Note that $\Sigma$ is a periodic quantity with period $Nm=\frac{2\pi}{L}$ 
since it is shifted by the transformation \eqref{eq:equivalence} 
with non-zero winding number $w~(V(x_2)=e^{2\pi i w \frac{x_2}{L}})$ 
as $\Sigma \rightarrow \Sigma + wNm$.}
\label{fracI}
\end{center}
\end{figure}

If we combine $N$ different types of fractional instantons, 
we obtain an instanton solution with $Q=1$ 
\begin{equation}
h_{{\cal I}\cdots{\cal I}} ~=~ 
\left( 1 + a_N e^{Nmz} \,,\, 
a_1 \,,\, 
\cdots \,,\, 
a_{N-1} \right) 
{\rm diag} \left[ 1 \, , \, e^{mz} \, , \, \cdots \, , \, e^{(N-1)mz} \right].  
\label{eq:cpn_instanton}
\end{equation}
These solutions have kink-like behaviors 
which can be seen from the function $\Sigma(x_1)$ defined by 
\cite{Isozumi:2004jc}
\begin{equation}
\Sigma(x_1) 
= \frac{1}{L} \int_0^L dx_2 \, \frac{i}{2 |h|^2} 
\left( h \p_2 h^\dagger - \p_2 h h^\dagger \right)
=\frac{m}{L} \int_0^L dx_2 \, 
\frac{\sum_{k=1}^{N-1}k\varphi^k\bar\varphi^k}
{1+\sum_{j=1}^{N-1}\varphi^j\bar\varphi^j} .
\label{eq:sigma}
\end{equation}
Fig.\,\ref{fracI} shows the function $\Sigma(x_1)$ for the BPS configurations 
with fractional and integer topological charges. 

\begin{figure}[htbp]
\begin{center}
\includegraphics[width=75mm]{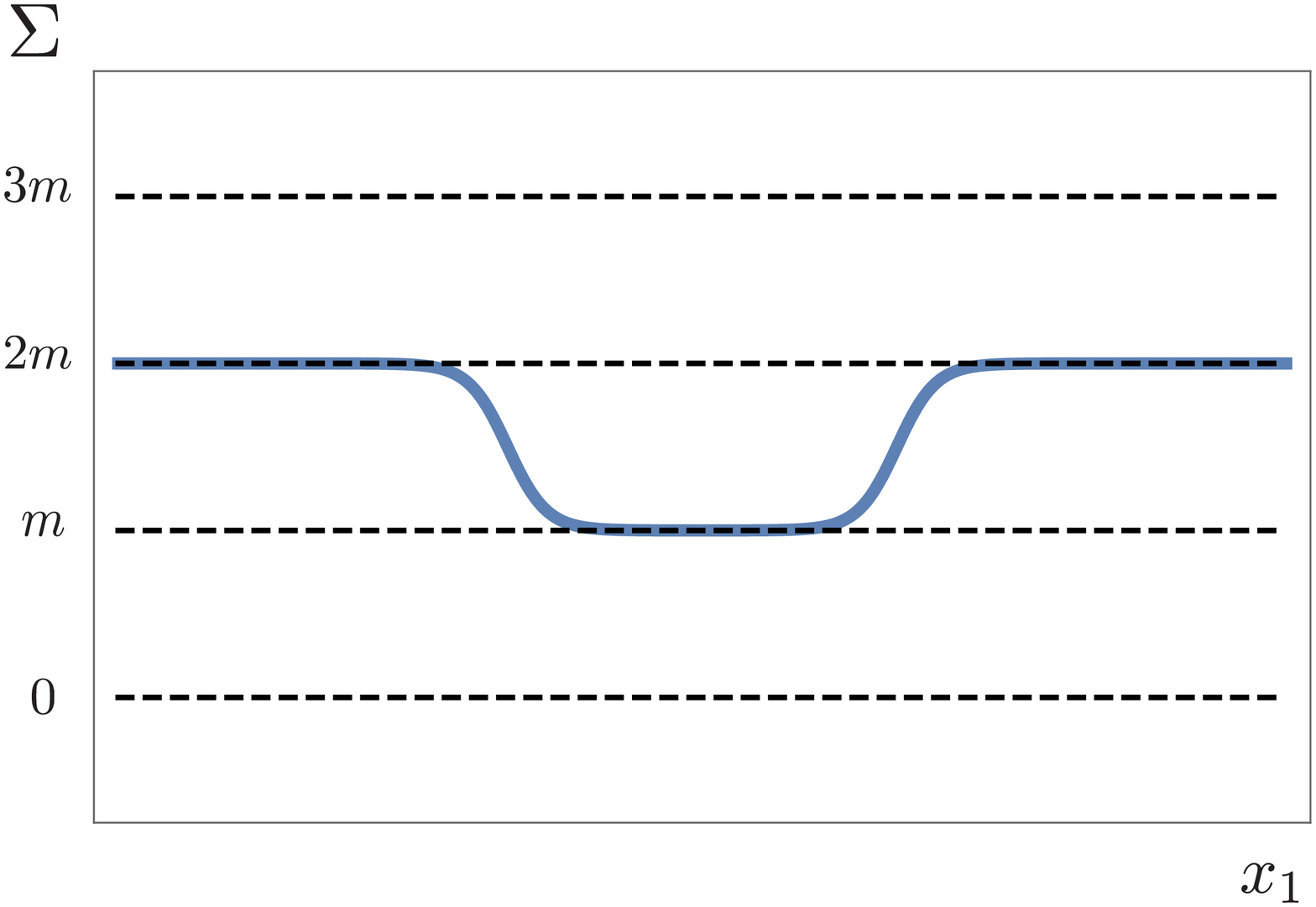}
\vs{-10}
\end{center}
\caption{Bion configuration is in ${\mathbb C}P^3$ model
with ${\mathbb Z}_{4}$-twisted boundary conditions.}
\label{bion}
\end{figure}

If we add a fractional anti-instanton to the left of a fractional instanton, 
we obtain a bion configuration, which has vanishing topological charge 
\cite{Misumi:2014jua}
\begin{equation}
h_{{\cal I}\bar{\cal I}} \, = \,
(0 \,,\, \cdots \,,\, 1 \,,\, 
a_+ e^{mx_1} + a_- e^{-mx_1} \,,\,
\cdots \,,\, 0) \ {\rm diag} \left[ 1 \, , \, e^{mx_2} \, , \, \cdots \, , \, e^{(N-1)mx_2} \right]. 
\label{eq:bion_cpn}
\end{equation}
The parameters $a_\pm$ are related to 
the positions and phases of the constituent fractional instantons $(x_\pm, \phi_\pm)$ as
\beq
a_\pm = \exp \left( \mp m x_\pm + i \phi_\pm \right). 
\eeq
Fig.~\ref{bion} illustrates the bion configuration. 
The bion configuration becomes a solution of field equation 
asymptotically for large separation $x_+-x_-$, 
and provides a basis to compute nonperturbative contributions relevant to the resurgence theory. 



\subsection{${\mathbb C}P^{N-1}$ quantum mechanics }

The two-dimensional field theory on ${\mathbb R} \times S^1$ is 
faithfully reproduced by the Kaluza-Klein decomposition 
\begin{equation}
\varphi^k(x_1,x_2) ~=~ \sum_{n \in \Z} \varphi^k_{(n)}(x_1) \,
\exp \left[ i\frac{2\pi}{L} \left( n+\frac{k}{N}\right) x_2 \right],
\label{eq:KKdecomCPN-1}
\end{equation}
where we used the mode functions taking into account of the 
${\mathbb Z}_N$-twisted boundary condition \eqref{eq:ZNboundCod_inhomo}. 
Integrating over the compactified coordinate $x_2$, the 
two-dimensional Lagrangian can be rewritten equivalently as 
a coupled system of infinitely many quantum mechanical variables 
$\varphi^k_{(n)}(x_1)$ 
\begin{equation}
{\cal L}_{\rm KK}=\int_0^L dx_2 \, {\cal L}_{\rm 2d}, 
\end{equation}
and the coupling constant of the quantum mechanics $g_{\rm 1d}$ can be defined as 
\begin{equation}
\frac{1}{g_{\rm 1d}^2}=\frac{L}{g_{\rm 2d}^2}.
\label{eq:1d2dCoupl}
\end{equation}
This infinite component quantum mechanics is equivalent to the 
two-dimensional field theory including the necessity of renormalization, 
and consequent asymptotic freedom. 
However, we can retain only finite number of quantum mechanical 
degrees of freedom and discard the infinite tower of Kaluza-Klein 
components, as long as we are interested in the semi-classical 
results from saddle points needed to catch the resurgence 
structure of the theory at low energies $(E \ll m$). 
Let us retain only the lowest mode $n=0$ for each $\varphi^k(x_1,x_2)$ 
out of an infinite tower of Kaluza-Klein variables to find the 
${\mathbb C}P^{N-1}$ quantum mechanics 
\begin{eqnarray}
L_{\rm 1d} = \frac{1}{g_{\rm 1d}^2} G_{k \bar l}^{(0)} 
\left[ \p_{x_1} \varphi_{(0)}^k \p_{x_1} \bar \varphi_{(0)}^l + 
k l m^2 \varphi_{(0)}^k \bar \varphi_{(0)}^l \right], 
\hs{10}
G_{k \bar l}^{(0)} = G_{k \bar l}(\varphi^i = \varphi_{(0)}^i).
\label{eq:cpN-qm}
\end{eqnarray}
We can now recognize that there are only $N$ discrete vacua 
corresponding to $h =$ $(1,0,\cdots,0)$, $\cdots$, $(0,\cdots,0,1)$ 
because of the ${\mathbb Z}_N$-twisted boundary condition. 

The ${\mathbb C}P^{N-1}$ quantum mechanics captures 
restricted configurations in the two-dimensional theory of the form 
\begin{equation}
h(x_1,x_2) \big|_{\rm QM} ~=~ h_{(0)}(x_1) \ {\rm diag}. \left[ 1 \, , \, e^{i m x_2} \, , \, \cdots \, , \, e^{i(N-1)m x_2} \right].
\label{eq:qm_conf}
\end{equation}
The function $\Sigma(x_1)$ exhibiting the kink-like profile in 
Eq.(\ref{eq:sigma}) 
reduces to 
\begin{equation}
\Sigma(x_1) 
=\frac{m\sum_{k=1}^{N-1}k\varphi^k_{(0)}\bar\varphi^k_{(0)}}
{1+\sum_{j=1}^{N-1}\varphi^j_{(0)}\bar\varphi^j_{(0)}} \; .
\label{eq:sigma_qm}
\end{equation}
in the  ${\mathbb C}P^{N-1}$ quantum mechanics. 
The fractional instanton in Eq.(\ref{eq:fractional_instanton}) 
and the bion configuration in Eq.\,(\ref{eq:bion_cpn}) 
take the form of Eq.\,(\ref{eq:qm_conf}).
Thus we find that both the fractional instantons and 
the bion configurations are correctly described 
in the ${\mathbb C}P^{N-1}$ quantum mechanics. 
We can show that all other multi-fractional instanton configurations 
can be correctly described by the ${\mathbb C}P^{N-1}$ quantum mechanics, 
provided it does not contain multi-fractional-instantons 
with $|Q| \ge 1$ anywhere locally. 

The instanton configuration (\ref{eq:cpn_instanton}) with $Q=1$ 
is not reducible to the ${\mathbb C}P^{N-1}$ quantum mechanics, 
since it does not satisfy (\ref{eq:qm_conf}). 
The action and topological charge densities 
of the $N$ fractional instanton solution in Eq.(\ref{eq:cpn_instanton}) 
do exhibit a strong $x_2$ dependence approaching the ordinary 
single instanton solution when the constituent fractional 
instantons are compressed in a point. 
This situation inevitably occurs whenever configurations with $|Q| \ge 1$ are contained. 
On the other hand, we find that the configurations compatible with the 
${\mathbb C}P^{N-1}$ quantum mechanics have action density and 
topological charge density which are independent of the 
coordinate $x_2$ of the compact direction. 
Therefore, configurations with $|Q|<1$ in the two-dimensional field theory 
are correctly captured by the ${\mathbb C}P^{N-1}$ quantum mechanics, 
provided the multi-fractional-instanton configurations with more than 
unit topological charge is not contained anywhere locally \cite{Misumi:2016fno}. 

Once it was conjectured that the ${\mathbb C}P^{N-1}$ model reduces 
to the sine-Gordon quantum mechanics in the limit of $L\to 0$ 
(the compactification limit) \cite{Dunne:2012ae,Dunne:2012zk}. 
However, it has been observed that the relative phase moduli 
of fractional instanton and anti-instanton is not correctly 
described by the sine-Gordon quantum mechanics \cite{Misumi:2014jua,Misumi:2014rsa,Misumi:2015dua}. 
We discuss the differences between 
the $\C P^1$ and the sine-Gordon quantum mechanics in Sec.\,\ref{sec:SG}.


\subsection{$\C P^1$ quantum mechanics with fermion and supersymmetry} 
To examine bion configurations, 
it is convenient to introduce a fermionic degree of freedom. 
Only in this subsection, we use Lorentzian signature instead of 
Euclidean signature in order to use also Schr\"odinger equation later. 
To denote 1d quantities simply, we rewrite without subscript: 
$-ix_1$ as the Lorentzian time $t$, 
$\varphi_{(0)}^{k=1} \rightarrow \varphi$, $g_{1d} \rightarrow g$, 
$G^{(0)}_{1\bar 1}\to G$ etc. 
The Lagrangian of the $\C P^1$ Lorentzian quantum 
mechanics with a fermion takes the form
\beq
L ~=~ \frac{1}{g^2} G \Big[ \p_t \varphi\p_t \bar\varphi 
- m^2 \varphi\bar\varphi 
+ i \bar \psi \D_t \psi + \epsilon m (1+ \varphi \, 
\p_\varphi \log G) \bar \psi \psi \Big], 
\label{eq:L_QM}
\eeq 
where $G$ is the Fubini-Study metric and $\D_t$ is the 
pullback of the covariant derivative
\beq
G = \frac{1}{(1+\varphi\bar\varphi)^2}, \hs{10} 
\D_t \psi = \Big[ \p_t + \p_t \varphi  \, 
\p_\varphi  \log G \Big] \psi. 
\eeq
The parameter $\epsilon$ controls the strength of the interaction 
between the bosonic and fermionic degrees of freedom. 
If we set $\epsilon = 1$, this model becomes a supersymmetric system 
which can be obtained from the 2d $\mathcal N=(2,0)$ $\C P^1$ 
sigma model by an analogous dimensional reduction 
as the one discussed in the previous subsection.
 
\begin{figure}[h]
\centering
\vs{-5}
\includegraphics[width=100mm]{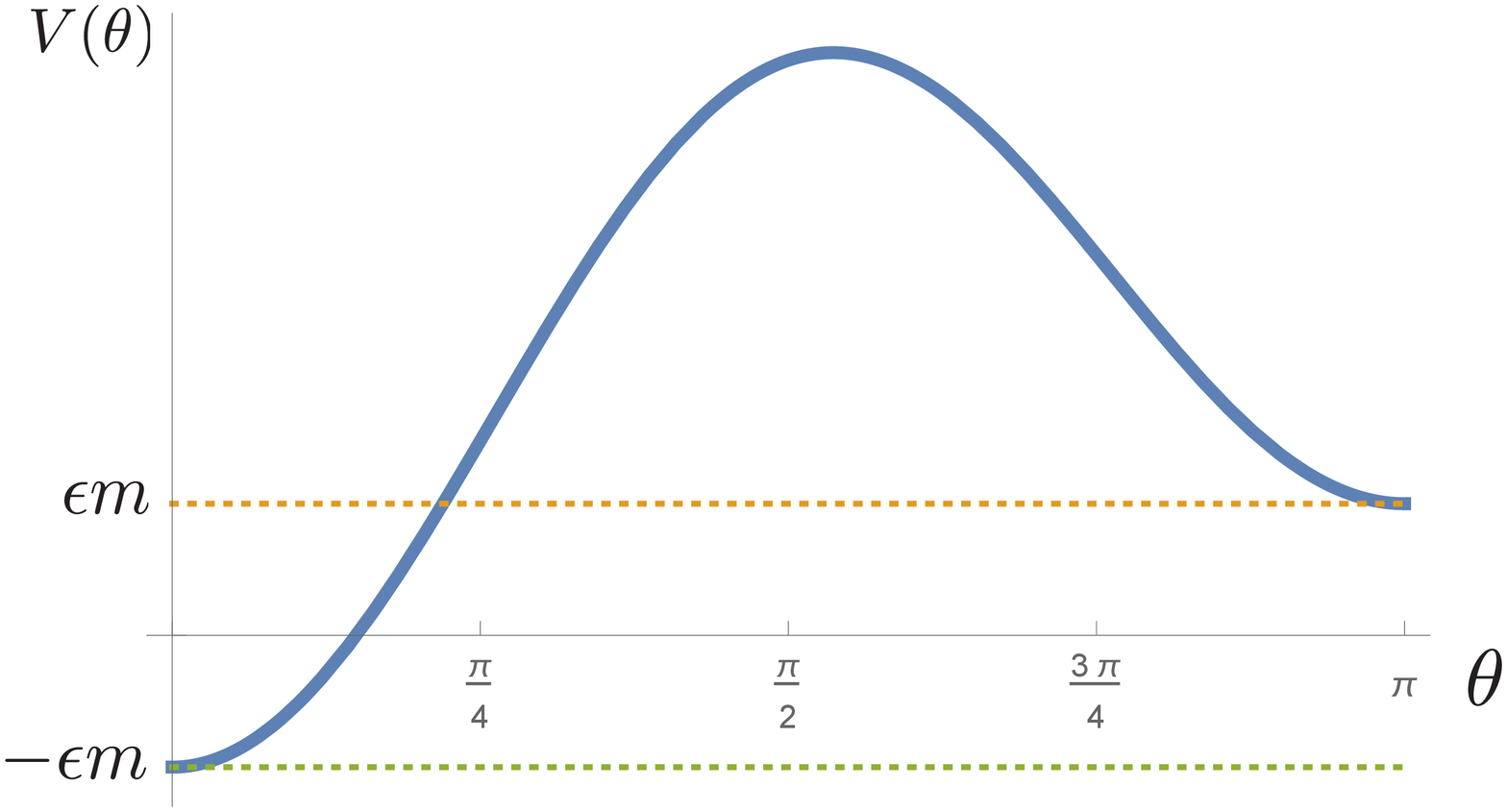} 
\vs{-10}
\caption{The potential $V$ with the contribution of the fermion. 
The horizontal axis denotes the latitude $\theta \equiv 2 {\rm arctan} |\varphi|$ on $\C P^1 \cong S^2$.}
\label{fig:potential}
\end{figure}

Since the fermion number $\bar \psi \psi$ commutes with the Hamiltonian, 
we can eliminate $\psi$ by using the conserved fermion number 
and the associated induced potential. 
By projecting quantum states onto the subspace of the Hilbert space 
with a fixed fermion number,
we obtain the following purely bosonic Lagrangian (see Appendix \ref{appendix:fermion} for details)
\beq
L ~=~ \frac{1}{g^2} \frac{\p_t \varphi\p_t\bar\varphi}
{(1+\varphi\bar\varphi)^2} 
- V(\varphi\bar\varphi),
\hs{10}
V(\varphi\bar\varphi) ~\equiv~ 
\frac{1}{g^2} \frac{m^2 \varphi\bar\varphi}
{(1+\varphi\bar\varphi)^2} - \epsilon m \frac{1-\varphi\bar\varphi}
{1+\varphi\bar\varphi}, 
\label{eq:Lagrangian}
\eeq
where we have chosen the fermion number 
so that the supersymmetric ground state for $\epsilon =1$ 
is contained in the subspace of the Hilbert space. 
The potential $V$ as a function of the latitude $\theta \equiv 2 {\rm arctan} |\varphi|$ is shown in Fig.\,\ref{fig:potential}.

For $\epsilon = 1$, the ground state wave function $\Psi_0$, 
which preserves the supersymmetry, 
is given as the zero energy solution of the Schr\"odinger equation 
\beq
H \Psi_0 = 0,
\eeq
with the Hamiltonian $H$ of the bosonic theory:
\beq
H = - g^2 (1+\varphi\bar\varphi)^2 \frac{\p}{\p \varphi} 
\frac{\p}{\p \bar \varphi} + V(\varphi\bar\varphi). 
\label{eq:H}
\eeq
We find the exact solution of the ground state wave function
\beq
\Psi_0 = \exp \left( \frac{m}{2g^2} \frac{1-\varphi\bar\varphi}
{1+\varphi\bar\varphi} \right). 
\label{eq:ground_state}
\eeq
The existence of the supersymmetric state implies that 
the ground state energy receives no non-perturbative correction 
for $\epsilon = 1$. 
For a generic value of $\epsilon$, 
there can be corrections to the ground state energy. 
Indeed, we can show that there exist non-perturbative corrections 
in the near supersymmetric case $\epsilon \approx 1$ 
by expanding the energy with respect to small $\delta \epsilon \equiv \epsilon-1$
\beq
E ~\approx~ \frac{\langle 0 |\delta H|0 \rangle}{\langle 0 |0 \rangle} ~=~ -\delta \epsilon m \left\langle \frac{1-\varphi\bar\varphi}{1+\varphi\bar\varphi} \right\rangle_{\epsilon=1},
\label{eq:nearsusy0}
\eeq
where the perturbative Hamiltonian is given by
\beq
\delta H ~=~ H - H_{\epsilon=1} ~=~ -\delta \epsilon m \frac{1-\varphi\bar\varphi}{1+\varphi\bar\varphi} . 
\eeq
As Eq.\,\eqref{eq:nearsusy0} indicates, 
we can exactly calculate the leading order coefficients 
in the small $\delta \epsilon$ expansion of the ground state energy 
by using the explicit form of the ground state wave function \eqref{eq:ground_state} as
\beq
E ~=~ \frac{\displaystyle \int dv \, \delta H |\Psi_0|^2}{\displaystyle \int dv \, |\Psi_0|^2} + \mathcal O( \delta \epsilon^2) ~=  \delta \epsilon \left[ g^2 - m \coth \frac{m}{g^2} \right] +  \mathcal O( \delta \epsilon^2), 
\label{eq:nearsusy}
\eeq
where $dv$ is the standard volume element on $\C P^1$: $dv \equiv d^2 \varphi/(1+\varphi\bar\varphi)^{2}$. 
Note that although we have expanded the energy with respect to $\delta \epsilon$, 
Eq.\,\eqref{eq:nearsusy} is non-perturbative as a function of the coupling constant $g$. 
We can decompose the ground state energy \eqref{eq:nearsusy} 
into the perturbative and non-perturbative parts
\beq
E_{\rm pert} &=& (g^2 -m) \delta \epsilon + \mathcal O\left( \delta \epsilon^2 \right), \phantom{\bigg[} \\
E_{\rm bion} &=& -2 m \, e^{- \frac{2m}{g^2}} \delta \epsilon + \mathcal O \left( e^{-\frac{4m}{g^2}} , \, \delta \epsilon^2 \right).
\label{eq:nearSUSY}
\eeq
It is interesting to note that perturbative 
contributions to this order of $\epsilon-1$ are terminated 
at the order $g^2$ without any higher order corrections, 
and hence there is no ambiguity associated with non-Borel summable asymptotic series.
In the following, we will see that contributions of (real and complex) bion 
configurations correctly reproduce this non-perturbative correction.


\section{Bion saddle points and one-loop approximation}\label{sec:oneloop}
In the previous section, we have seen that 
the ground state energy receives no correction for $\epsilon=1$ due to the supersymmetry 
and there exists a non-perturbative correction 
at least in the near supersymmetric case $\epsilon \approx 1$. 
In this section, we discuss the non-perturbative correction 
from the viewpoint of the saddle point method 
in the projected model \eqref{eq:Lagrangian},
which we complexify to find a complex bion solution.  

In the path integral formalism, 
the ground state energy can be obtained from the partition function
\beq
Z(\beta) = \int \D \varphi \, \exp(- S_E[\varphi] ),
\eeq
where the path integral is performed 
over configurations satisfying the periodic boundary condition 
$\varphi(\tau+\beta) = \varphi(\tau)$. 
The asymptotic behavior of the partition function 
in the weak coupling limit $g \rightarrow 0$ 
can be obtained by using the saddle point method in the Gaussian approximation:
\beq
Z = \sum_{\sigma \in \mathfrak S} e^{-S_\sigma} \Big[ ( \det \Delta_\sigma )^{-\frac{1}{2}} + \mathcal O(g) \Big],
\eeq 
where $\mathfrak S$ denotes a 
set of saddle points of $S_E$, 
$S_\sigma$ is the value of the action at the saddle point $\sigma$ 
and $\det \Delta_\sigma$ is the one-loop determinant in the saddle point configuration.
Let $Z_0$ be the contributions from the trivial saddle point (classical vacuum) and 
$Z_1$ be the leading order correction from non-trivial saddle points: 
\beq
Z = Z_0 + Z_1 + \cdots. 
\eeq
Then the non-perturbative correction to the ground state energy 
can be obtained as
\beq
E ~=~ - \lim_{\beta \rightarrow \infty} \frac{1}{\beta} \log Z 
~\approx~ - \lim_{\beta \rightarrow \infty} \frac{1}{\beta} \left[ \log Z_0 + \frac{Z_1}{Z_0} + \cdots \right]. 
\label{eq:energy}
\eeq
In the following, we compute $Z_1/Z_0$ by finding saddle points and 
calculating one-loop determinants around the saddle points.

\subsection{Real bion solution}
Let us look for saddle points of the Euclidean action
which are responsible for the non-perturbative correction
by solving the classical equation of motion. 
Since we are interested in the zero temperature limit, 
we consider saddle points in the limit $\beta \rightarrow \infty$. 
In this subsection, as the first step, we show that there is a real bion solution 
in the model with the potential modified by the fermionic degree of freedom. 

To find out classical solutions, 
it is convenient to use symmetries of the Euclidean action 
and associated conservation laws. 
Since the Euclidean action
\beq
S_E ~= \, \int d\tau \left[ \frac{1}{g^2} 
\frac{\p_\tau \varphi\p_\tau \bar \varphi}{(1+\varphi\bar\varphi)^2} 
+ V(\varphi\bar\varphi) \right],
\eeq
is invariant under the shift of the Euclidean time $\tau \rightarrow \tau - \tau_0$, 
the corresponding ``energy" is a conserved quantity
\beq
E ~\equiv~ \frac{1}{g^2} \frac{\p_\tau \varphi\p_\tau\bar\varphi }
{(1+\varphi\bar\varphi)^2} - V(\varphi\bar\varphi).
\eeq
The action is also invariant under the phase rotation 
$\varphi \rightarrow e^{i \phi} \varphi$, 
so that the corresponding angular momentum is a conserved charge
\beq
l ~\equiv~ \frac{i}{g^2} \frac{\p_\tau \varphi \bar \varphi 
- \p_\tau \bar \varphi \varphi}{\, (1+\varphi\bar\varphi)^2}. 
\eeq
Since we are interested in saddle point configurations with finite action, 
we impose the boundary condition 
so that $\varphi$ is at the minimum of the potential for 
$\tau \rightarrow \pm \infty$: 
\beq
\lim_{\tau \rightarrow \pm \infty} \varphi 
= \lim_{\tau \rightarrow \pm \infty} \bar \varphi = 0.
\label{eq:BC}
\eeq 
Then it follows that the saddle point configuration 
cannot have the angular momentum $l$, 
i.e. the phase of $\varphi$ is a constant of motion.  
In addition, ``the energy conservation law" implies that
\beq
\frac{1}{g^2} \frac{\p_\tau \varphi\p_\tau\bar\varphi}
{(1+\varphi\bar\varphi)^2} 
- V(\varphi\bar\varphi) ~=~ \epsilon m ~=~ E |_{\varphi=0} .
\eeq
We can integrate the energy conservation law to obtain 
``the real bion solution".
\beq
\varphi ~=~ e^{i \phi_0} \sqrt{\frac{\omega^2}{\omega^2 - m^2}} 
\frac{1}{i \sinh \omega (\tau-\tau_0)}, 
\label{eq:real_bion}
\eeq
where $\omega$ is given by
\beq
\omega \equiv m \sqrt{1 + \frac{2 \epsilon g^2}{m}}. 
\eeq
The parameters $\tau_0$ and $\phi_0$ are integration constants, 
i.e. moduli parameters.
The orbit of this solution in $\C P^1$ is 
a great circle starting from the south pole 
($\varphi=0$, minimum of the potential) 
and passing through the north pole at $\tau = \tau_0$. 
The phase modulus $\phi_0$ is related to 
the longitude of the great circle. 
Thus the moduli space of real bion is a cylinder 
\beq
\mathcal M_{\rm bion} = \R \times S^1.
\eeq  
These parameters will be eventually integrated to incorporate the contribution of all the bion solutions
to the partition function. Precisely speaking, the first factor should be interpreted 
as an infinitely large $S^1$ with radius $\beta \rightarrow \infty$, 
along which the moduli integration gives a factor of $\beta$ to the single bion contribution $Z_1$.

The real bion solution can be viewed as a kink-antikink solution 
with fixed relative position and phase,
since it can be rewritten into the kink-antikink form 
\eqref{eq:bion_cpn} 
in terms of the homogeneous coordinates:
\beq
h &=& \left( \, 1 \, , a_+ e^{\omega \tau} + a_- e^{-\omega \tau} \, 
\, \right), \hs{10} 
a_+ = e^{ -\omega \tau_+ - i \phi_+}, \hs{5}
a_- = e^{\omega \tau_- - i \phi_-}, 
\label{eq:ansatz}
\eeq
where the positions and phases are given by
\beq
\tau_\pm &=& \tau_0 \pm \frac{1}{2\omega} 
\log \frac{4\omega^2}{\omega^2-m^2}, \hs{10}
\phi_\pm ~=~ \phi_0 \mp \frac{\pi}{2}.
\label{eq:rb_moduli}
\eeq

The Lagrangian for this saddle point configuration takes the form
\beq
L ~=~ 4 m \epsilon \Big[ f(\tau-\tau_0) 
\cosh \omega (\tau-\tau_0) \Big]^2 - m \epsilon, 
\label{eq:lagrangian_solution}
\eeq
where the function $f(\tau)$ is given by 
\beq
f(\tau) \equiv \frac{\omega^2}{\omega^2 + 
(\omega^2-m^2) \sinh^2 \omega \tau}. 
\label{eq:function}
\eeq
Neglecting the vacuum value of the Lagrangian, 
we obtain the action for the real bion as
\beq
S_{\rm rb} = 4 m \epsilon \int_{-\infty}^{\infty} d \tau 
\Big[ f(\tau-\tau_0) \cosh \omega (\tau-\tau_0) \Big]^2 
= \frac{2\omega}{g^2} + 2 \epsilon \log \frac{\omega+m}{\omega-m}.
\eeq
This implies that the real bion can give a non-perturbative correction 
of order $e^{-S_{\rm rb}} \sim e^{-\frac{2\omega}{g^2}}$. 
However, as we have seen in the previous section, 
the ground state energy does not receive any correction for 
$\epsilon=1$,
and hence there should be other saddle point configurations
which cancel the contribution of the real bion solution.

\subsection{Complex bion solution}\label{subsec:complexbion}
The absence of non-perturbative correction at $\epsilon=1$ implies that 
there are other saddle points which should be taken into account. 
However, the configuration \eqref{eq:real_bion} is the general solution 
satisfying the boundary condition \eqref{eq:BC}. 
The only way to obtain other saddle point configurations is 
to extend the configuration space by complexifying the degrees of freedom. 
Such a procedure is a straightforward generalization of 
the complexification of integration contour 
for ordinary finite dimensional integrals,
which is a necessary step in the saddle point method \cite{Behtash:2015zha}. 

In the case of the complex field $\varphi$, 
we independently complexify its real and imaginary parts: 
\beq
(\varphi, \bar \varphi) = (\varphi_R + i \varphi_I \, , \, \varphi_R - i \varphi_I) ~~~\longrightarrow~~~ 
(\varphi_R^\C + i \varphi_I^\C \, , \, \varphi_R^\C - i \varphi_I^\C). 
\eeq
Consequently, $\bar \varphi$ becomes an independent complex 
degree of freedom which is not related to $\varphi$ by complex conjugation.  
In the following, we denote $\tilde \varphi$ 
for the complexification of $\bar \varphi$ 
to avoid confusion
\beq
\bar \varphi ~ \rightarrow~ \tilde \varphi ~\not = \, 
\mbox{ complex conjugate of $\varphi$}.
\eeq
Then we regard the action $S[\varphi,\tilde \varphi]$ 
as an analytically continued holomorphic functional of the complexified degrees of freedom 
\beq
S
[\varphi,\tilde \varphi] ~= \, \int d\tau \left[ \frac{1}{g^2} 
\frac{\p_\tau \varphi\p_\tau \tilde \varphi}{(1+\varphi\tilde\varphi)^2} 
+ V(\varphi\tilde\varphi) \right].
\label{eq:complexified_cp1}
\eeq
We also impose the boundary condition (\ref{eq:BC}) with 
$\tilde \varphi$ replacing $\bar \varphi$. 
By deforming integration contour, 
the integral can be expressed as a sum of contributions from 
a 
set of saddle points of the complexified action $S[\varphi,\tilde \varphi]$.

Since the action is extended as a holomorphic functional, 
it is invariant under the symmetries of the original action 
with complexified transformation parameters. 
Furthermore, the complexified equations of motion 
take the same forms as those of the original action
and hence the configuration \eqref{eq:real_bion} is still the general solution 
satisfying the boundary condition \eqref{eq:BC}. 
The important difference in the complexified case is that 
the integration constants $\tau_0$ and $\phi_0$ are now complex parameters,
i.e. the solution $(\varphi, \tilde \varphi)$ is 
a holomorphic function of the moduli parameters $\tau_0$ and $\phi_0$.

The solution \eqref{eq:real_bion} smoothly varies 
under small shifts of moduli parameters $\tau_0$ and $\phi_0$. 
Such solutions are simply related to the real bion solution 
by the complexified symmetry transformations, 
and the value of the corresponding action remains the same. 
Thus the moduli space of bion configurations is also complexified:
$\mathcal M_{\rm bion} \rightarrow \mathcal M_{\rm bion}^\C$. 
The integration contour of the moduli integral for the partition function 
can be any middle dimensional contour 
in the complexifed moduli space $\mathcal M_{\rm bion}^\C$
as long as it is related to the original real contour $\mathcal M_{\rm bion}$ 
by a continuous deformation. 
However, for our purpose, 
we do not need to consider the deformation of the integration contour
and the moduli integral will be performed over 
$\mathcal M_{\rm bion}$ in the next section. 
Thus, the bion configurations with the complexified moduli do not give 
a physically distinct contribution, 
until the shift in the complexified transformation meets a singularity 
and produces a jump in the value of the action.

The singular solution can be obtained by a shift, for instance by an amount
\beq
\tau_0 ~\rightarrow~ \tilde \tau_0 = \tau_0 + \frac{1}{\omega} \frac{\pi i}{2},
\eeq 
under which the solution becomes
\beq
\varphi = e^{i \phi_0} \sqrt{\frac{\omega^2}{\omega^2-m^2}} \frac{1}{\cosh \omega(\tau - \tau_0)}, \hs{7}
\tilde \varphi = -e^{-i \phi_0} \sqrt{\frac{\omega^2}{\omega^2-m^2}} \frac{1}{\cosh \omega(\tau - \tau_0)}.
\eeq 
As we will see below, this configuration has singularities 
at which the action density diverges. 
Since $\tilde \varphi$ is no longer the complex conjugate of $\varphi$, 
this is a solution of the complexified model 
and hence we call this configuration ``complex bion solution". 

\begin{figure}
\begin{minipage}{0.45\hsize}
\centering
\includegraphics[width=75mm]{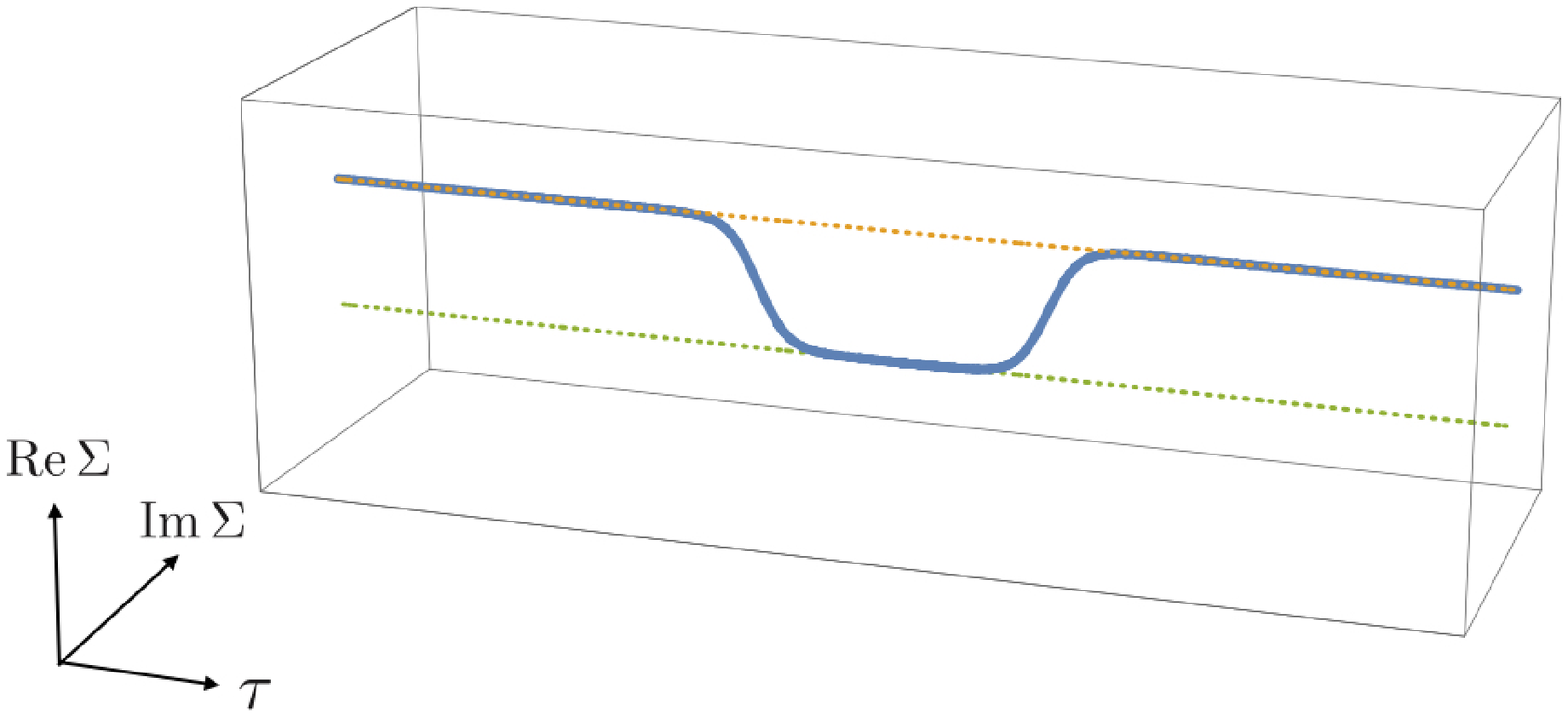} \\ 
(a) $\Sigma(\tau)$ for real bion
\end{minipage}
\hs{10}
\begin{minipage}{0.45\hsize}
\centering
\includegraphics[width=75mm]{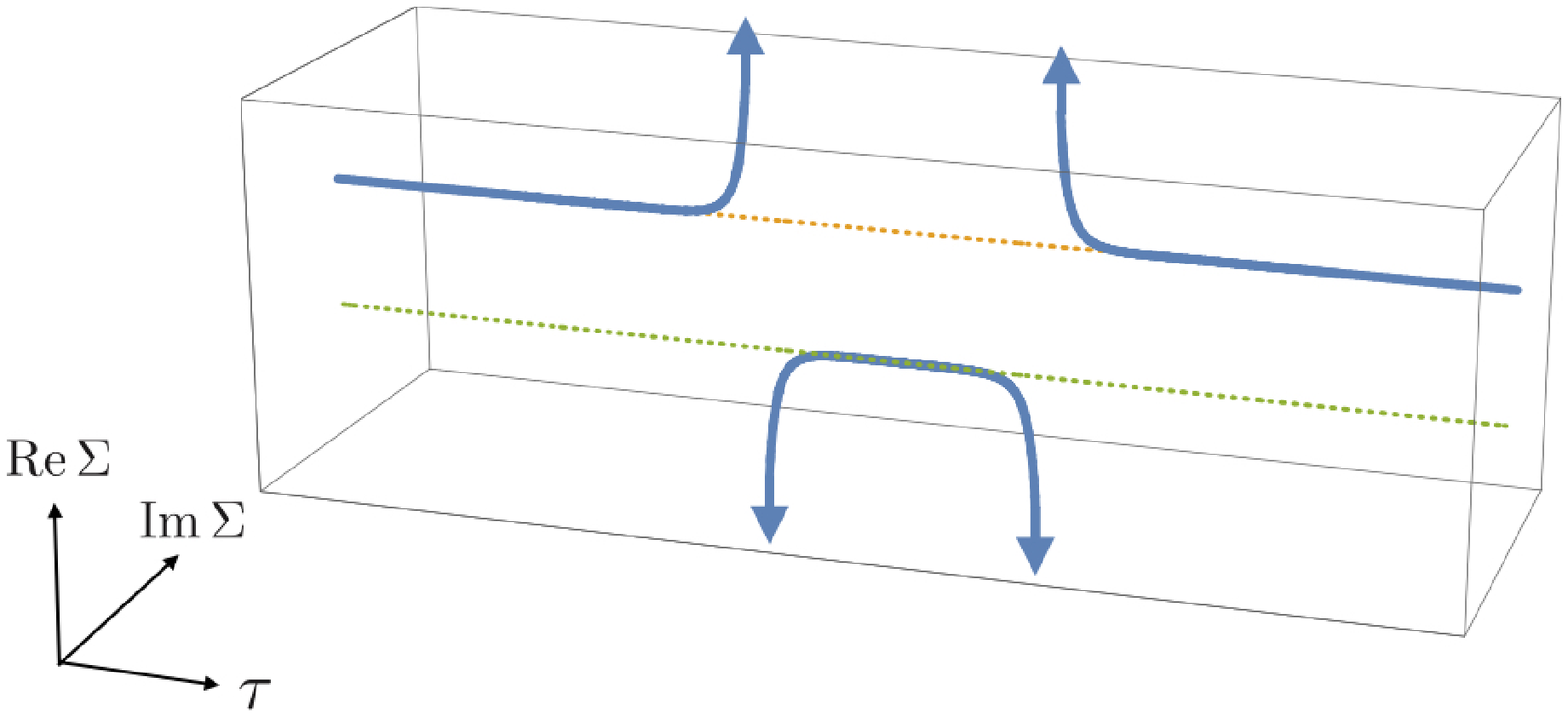} \\
(b) $\Sigma(\tau)$ for complex bion
\end{minipage}
\caption{
Kink profiles of for real and complex bions. 
The complex bion solution has singularities at which $\Sigma(\tau)$ diverges. 
Note that $\Sigma(\tau)$ can also be complex in the complexified model.}
\label{fig:bions}
\end{figure}

It is worth noting that the shifted solution can also be rewritten 
into the kink-antikink form 
\beq
\varphi = \left( e^{\omega (\tau - \tau_+) - i\phi_+} + e^{-\omega (\tau - \tau_-)-i\phi_-} \right)^{-1}, \hs{5}
\tilde \varphi = \left( e^{\omega (\tau - \tau_+)+i\phi_+} + e^{-\omega (\tau - \tau_-)+i\phi_-} \right)^{-1}, 
\eeq
with complexified position parameters $\tau_\pm$:
\beq
\tau_\pm &=& \tau_0 \pm \frac{1}{2\omega} 
\left( \log \frac{4\omega^2}{\omega^2-m^2} + \pi i \right)
, \hs{10} 
\phi_\pm = \phi_0 - \frac{\pi}{2},
\label{eq:cb_moduli}
\eeq
where we have used the fact that 
the shift $\tau_0 \rightarrow \tau_0 + \frac{1}{\omega} \frac{\pi i}{2}$ can be rewritten as
the combination of the shifts 
$\omega \tau_{+} \pm i \phi_{+} \rightarrow \omega \tau_{+} \pm i \phi_{+} + \frac{\pi i}{2} \, ({\rm mod} \, 2\pi i)$
and 
$\omega \tau_{-} \pm i \phi_{-} \rightarrow \omega \tau_{-} \pm i \phi_{-} + \frac{\pi i}{2} \, ({\rm mod} \, 2\pi i)$.
Therefore, the complex bion solution can also be viewed 
as a kink-antikink solution with complex relative distance
\beq
\tau_r ~\equiv~ \tau_+-\tau_- \, =~ \frac{1}{\omega} \left( \log \frac{4\omega^2}{\omega^2-m^2} + \pi i \right). 
\eeq 
Fig.\,\ref{fig:bions} shows
the kink-like profiles of the function $\Sigma(\tau)$ in Eq.(\ref{eq:sigma_qm}),
which takes the following form in the complexified theory
\beq
\Sigma = m \frac{\varphi \tilde \varphi}{1+\varphi \tilde \varphi}.
\eeq

\begin{figure}[h]
\centering
\includegraphics[width=80mm]{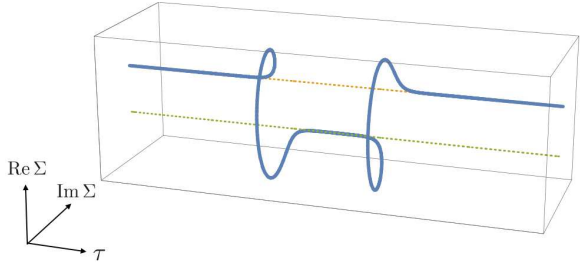}
\caption{Kink profile of regularized complex bion}
\label{fig:regularizedbion}
\end{figure}

\begin{figure}
\begin{minipage}{0.45\hsize}
\centering
\includegraphics[width=70mm]{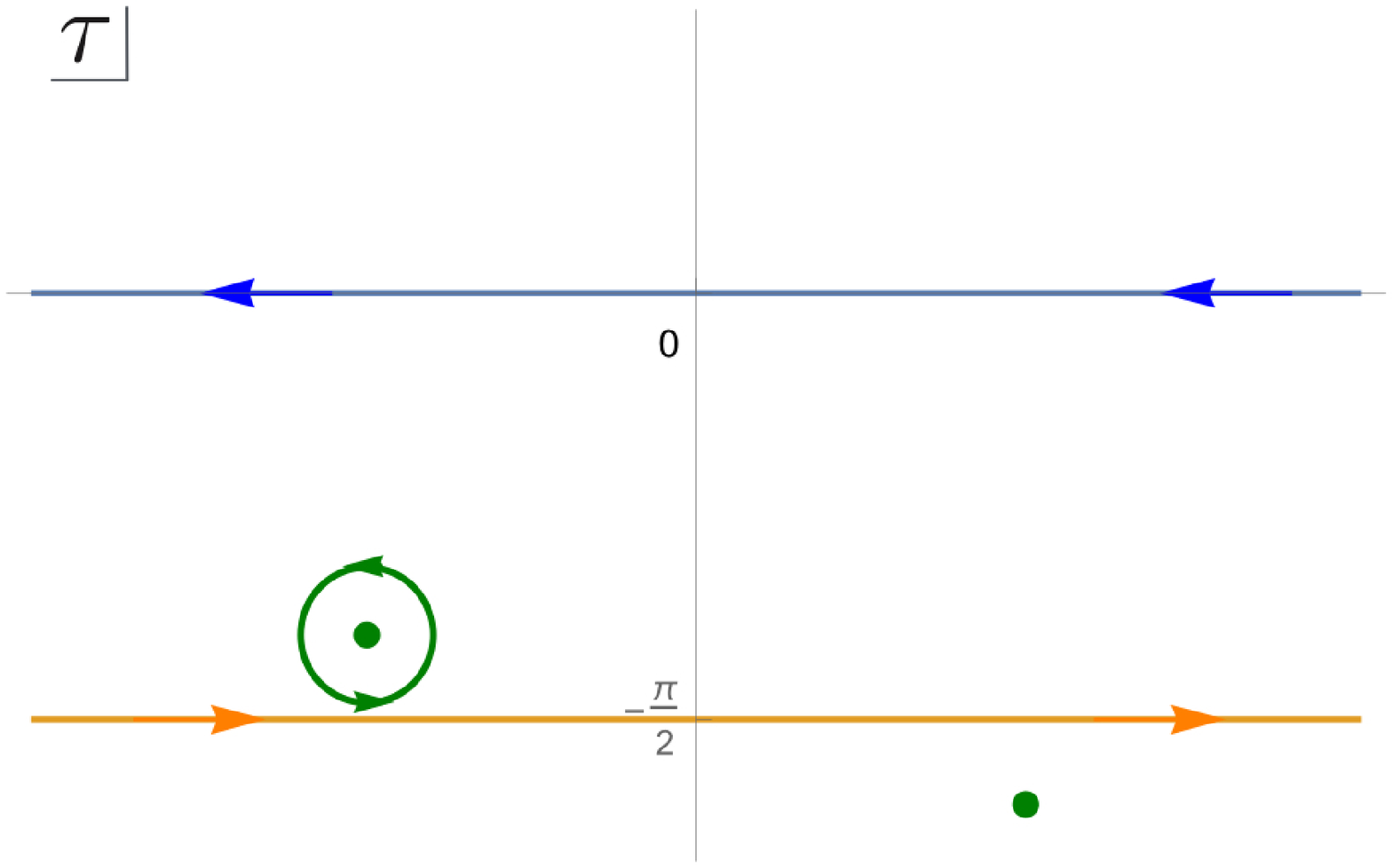} \\ 
(a) $\theta = {\rm arg} \, g^2 > 0$
\end{minipage}
\hs{10}
\begin{minipage}{0.45\hsize}
\centering
\includegraphics[width=70mm]{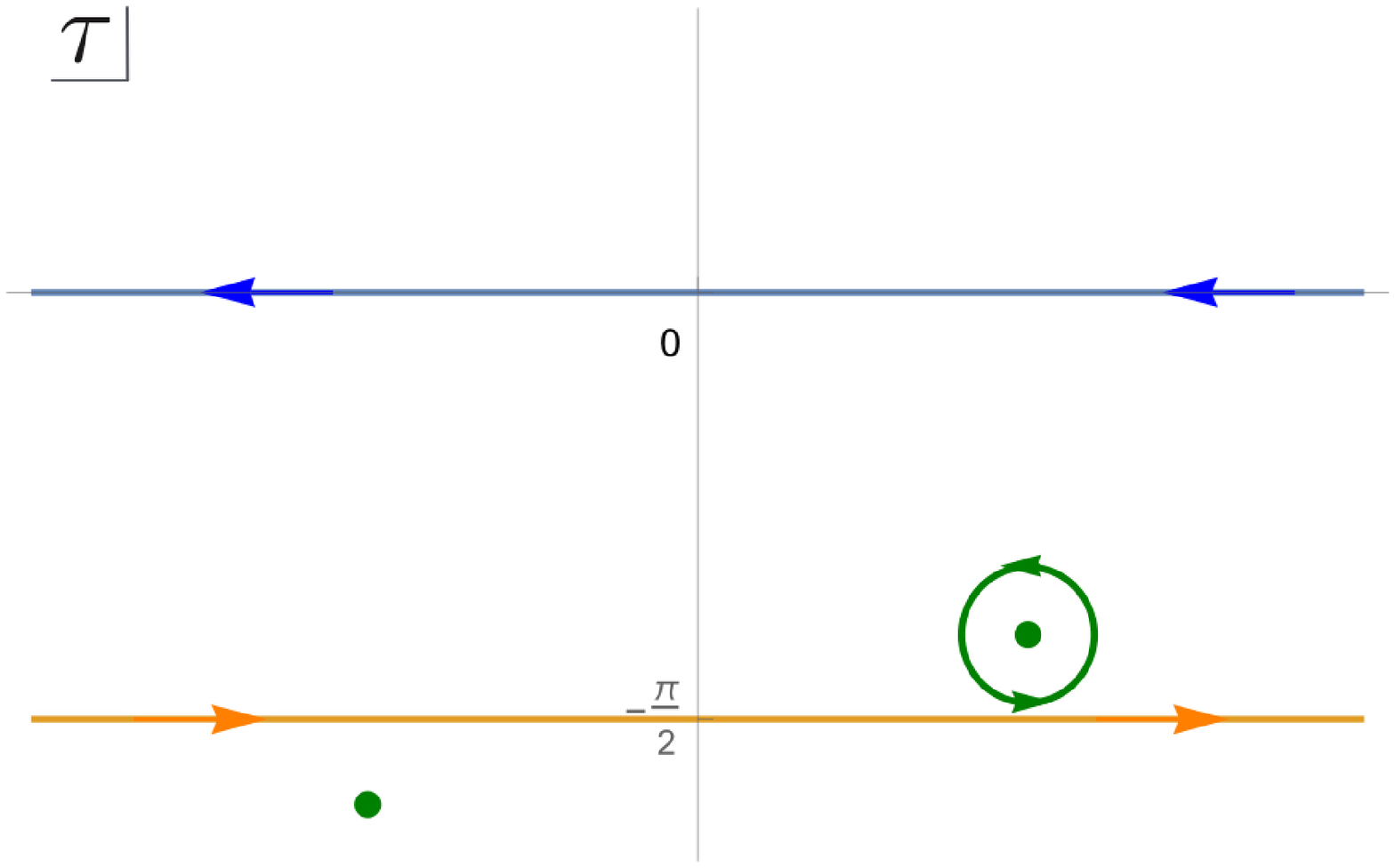} \\
(b) $\theta = {\rm arg} \, g^2 < 0$
\end{minipage}
\caption{The integration contour for $S_{\rm cb} - S_{\rm rb}$ in the complex $\tau$ plane. 
Depending on the sign of ${\rm arg} \, g^2$, 
the difference of the action $S_{\rm cb} - S_{\rm rb}$ is given by 
the residue at either $\tau_{\rm pole}^{+}$ or $\tau_{\rm pole}^{-}$.}
\label{fig:pole}
\end{figure}

The value of the 
Lagrangian in Eq.(\ref{eq:lagrangian_solution}) 
for the shifted solution ($\tau_0\to \tilde \tau_0$) 
is given in terms of the function $f$ defined in Eq.(\ref{eq:function}) as 
\begin{eqnarray}
L = 4 m \epsilon \Big[ f \left( \tau - \tilde \tau_0 \right) 
\cosh \omega(\tau - \tilde \tau_0) \Big]^2
= - 4 m \epsilon \left[ \frac{\omega^2 \sinh \omega(\tau-\tau_0)}
{\omega^2 - (\omega^2-m^2) \cosh^2 \omega(\tau-\tau_0)} \right]^2, 
\end{eqnarray}
where we have neglected the vacuum value of the Lagrangian. 
Since it has second order poles at 
\beq
\tau_{\rm pole}^{\pm} = \tau_0 \pm \frac{1}{\omega} {\rm arccosh} 
\sqrt{\frac{\omega^2}{\omega^2-m^2}}, 
\eeq
the shifted solution is a singular solution.
To regularize the acton, let us turn on small imaginary part of the coupling constant: $\theta \equiv \arg \, g^2$. 
Fig.\,\ref{fig:regularizedbion} shows the kink profile of the regularized complex bion. 
Neglecting the vacuum value of the Lagrangian, we obtain 
the action for the complex bion solution as
\beq
S_{\rm cb} = 4 m \epsilon \int_{-\infty}^{\infty} d\tau 
\Big[ f \left( \tau - \tilde \tau_0 \right) 
\cosh \omega(\tau - \tilde \tau_0) \Big]^2
= 4 m \epsilon \int_C d \tau \Big[ f \left( \tau - \tau_0 \right) \cosh \omega(\tau - \tau_0) \Big]^2.
\label{eq:Scb}
\eeq
The integrand in the last expression is the same as that for the real bion 
whereas the integration contour $C$ is 
the line ${\rm Im} \, \tau = - \frac{1}{\omega} \frac{\pi}{2}$ instead of the real axis, 
and hence the difference of $S_{\rm rb}$ and $S_{\rm cb}$ can be calculated by 
deforming the integration contour from the real axis to $C$. 
Although the action is invariant under any smooth deformation of the contour,
its value jumps when one of the poles crosses the contour. 
By evaluating the residue at the pole, 
we can show that the difference of the action for the real and complex bion is given by
\beq
S_{\rm cb} &=& S_{\rm rb} \pm 2\pi i \epsilon. 
\eeq
As shown in Fig.\,\ref{fig:pole}, 
the difference of the action $S_{\rm cb} - S_{\rm rb}$ is given by 
the residue at either $\tau_{\rm pole}^{+}$ or $\tau_{\rm pole}^{-}$ 
depending on the sign of ${\rm arg} \, g^2$. 
Thus, the contribution of the complex bion has an ambiguity for a generic value of $\epsilon$. 
In the supersymmetic case $\epsilon=1$, 
the difference $S_{\rm cb} - S_{\rm rb}$ is $2\pi i$,
and hence there is no ambiguity $\exp(-S_{\rm cb}) = \exp (-S_{\rm rb})$. 

From the periodicity of the solution and the Lagrangian under 
the shift of the imaginary part of $\tau_0$, 
we find that there are only two distinct classes of solutions : 
real and complex bion solutions as exact solutions of the complexified equation of motion 
in the ${\mathbb C}P^1$ qunatum mechanics. 
To see that the contributions of the real and complex bions cancel out for $\epsilon =1$, 
we have to evaluate the one-loop determinant in the bion background. 

\subsection{One-loop determinant in bion backgrounds}
In this section, 
we compute the contributions of the bion configurations 
by evaluating the one-loop determinant 
for the fluctuations around the bion backgrounds. 
For this purpose, we expand the action 
with respect to the fluctuations $\xi^a~(a=1,2)$ defined by
\beq
\ba{c} 
\varphi \\
\tilde \varphi 
\ea 
= 
\ba{c} 
\varphi_{\rm sol} \\
\tilde \varphi_{\rm sol} 
\ea 
+ g
\ba{cc}
e_1 & e_2 \\
\bar e_1 & \bar e_2 
\ea
\ba{c}
\xi^1 \\
\xi^2
\ea,
\eeq

where the background bion solution is given by
\beq
\varphi_{\rm sol} = e^{i \phi_0} \sqrt{\frac{\omega^2}{\omega^2-m^2}} \frac{1}{i \sinh \omega(\tau-\tau_0)}, \hs{5}
\tilde \varphi_{\rm sol} = -e^{-i \phi_0} \sqrt{\frac{\omega^2}{\omega^2-m^2}} \frac{1}{i \sinh \omega(\tau-\tau_0)}, 
\eeq
with 
\beq
{\rm Im} \, \tau_0 = 
\left\{ \begin{array}{ll} 0 & \mbox{for real bion} \\ \frac{\pi i}{2\omega} & \mbox{for complex bion} \end{array} \right.,
\label{eq:imaginarytau0}
\eeq
It is convenient to use the following basis of vielbein:
\beq
\ba{cc}
e_1 & e_2 \\
\bar e_1 & \bar e_2 
\ea
= 
(1+|\varphi_{\rm sol}|^2)
\ba{cc}
\frac{1}{\sqrt{2}} & \phantom{-} \frac{i}{\sqrt{2}} \\
\frac{1}{\sqrt{2}} & -\frac{i}{\sqrt{2}}
\ea.
\eeq
Then we obtain the quadratic action of the form
\beq
\delta S = \frac{1}{2} \int d\tau \, \xi^{\rm T} \Delta \, \xi, \hs{10}
\xi =
\ba{c}
\xi^1 \\
\xi^2
\ea,
\eeq
where $\Delta$ is a 2-by-2 matrix of second order differential 
operator around the bion background.
The contribution from each bion is given by the determinant of the operator $\Delta$:
\beq
\frac{Z_1}{Z_0} = \sum_{\rm bion} e^{-S} 
\frac{\int \D \xi \exp \left( - \frac{1}{2} \int d\tau \, \xi^{\rm T} \Delta \, \xi \right)}{\int \D \xi \exp \left( - \frac{1}{2} \int d\tau \, \xi^{\rm T} \Delta_0 \xi \right)}
= \sum_{\rm bion} e^{-S} \int d\tau_0 d \phi_0 \, \sqrt{\det \left( \frac{\mathcal G}{2\pi} \right)} \sqrt{\frac{\det \Delta_0}{\det' \! \Delta \ }},
\eeq
where $\Delta_0 = - \p_\tau^2 + \omega^2$ is the differential operator around the minimum of the potential
and ${\det}' \Delta$ denotes the determinant excluding the zero modes.
The contribution of zero modes are expressed as the moduli integral 
over the real moduli space $\mathcal M_{\rm bion} = \R \times S^1$
whose measure is given by the moduli space metric $\mathcal G$ (see Appendix \ref{appendix:measure}). 

To compute the determinant, it is convenient to introduce $\Xi_0$ defined as
\beq
\Xi_0 &\equiv&
\frac{1}{g} 
\ba{cc} 
e_1 & e_2 \\ 
\bar e_1 & \bar e_2 
\ea^{-1}
\ba{cc} 
\p_{\tau_0} \varphi_{\rm sol} & \p_{\phi_0} \varphi_{\rm sol} \\
\p_{\tau_0} \bar \varphi_{\rm sol} & \p_{\phi_0} \bar \varphi_{\rm sol} 
\ea.
\label{eq:Xi_0}
\eeq
Its explicit form is 
\beq
\Xi_0 &=& \frac{\sqrt{2}}{g} \frac{\sqrt{\omega^2-m^2}}{\omega} f(\tau-\tau_0)
\ba{cc}
\omega \cos \phi_0 \cosh \omega(\tau-\tau_0) & - \sin \phi_0 \sinh \omega(\tau-\tau_0) \\
\omega \sin \phi_0 \cosh \omega(\tau-\tau_0) & \phantom{-} \cos \phi_0 \sinh \omega(\tau-\tau_0)
\label{eq:explicitXi0}
\ea. 
\eeq
In terms of $\Xi_0$, the differential operator $\Delta$ can be simply written as
\beq
\Delta = - \left( \p_\tau + \p_\tau \Xi_0 \Xi_0^{-1} \right) \left( \p_\tau - \p_\tau \Xi_0 \Xi_0^{-1} \right).
\label{eq:Delta}
\eeq
It is clear from this form of $\Delta$ that 
the column vectors of $\Xi_0$ are zero modes of the operator $\Delta$, i.e. 
$\Xi_0$ is a basis of the zero modes. 
As shown in Appendix \ref{appendix:determinant}, 
the determinant of operators of the form \eqref{eq:Delta} 
can be obtained by using the formula
\beq
\frac{\det' \! \Delta \ }{\det \Delta_0} ~=~ \frac{\det \mathcal G}{\det \left( \, 2 \, M K_+^\dagger K_- \right)},
\label{eq:determinant}
\eeq
where the matrices $\mathcal G$, $M$ and $K_\pm$ are defined by
\beq
\mathcal G = \int_{-\infty}^\infty d \tau \, \Xi_0^\dagger \Xi_0, \hs{10} 
M = \mp \lim_{\tau \rightarrow \pm \infty} \p_\tau \Xi_0 \Xi_0^{-1}, \hs{10} 
K_\pm = \lim_{\tau \rightarrow \pm \infty} e^{\pm M \tau} \Xi_0. 
\eeq
In the present case, $M = \omega \mathbf 1_2$ and
$\mathcal G$ is the moduli space metric.  
The matrices $K_\pm$ can be calculated from the explicit form of $\Xi_0$ \eqref{eq:explicitXi0} as
\beq
K_\pm = \frac{2\sqrt{2}}{g} \frac{\omega}{\sqrt{\omega^2-m^2}} e^{\pm \omega \tau_0}
\ba{cc}
\omega \cos \phi_0 & \mp \sin \phi_0 \\
\omega \sin \phi_0 & \pm \cos \phi_0
\ea.
\eeq
Substituting $K_\pm$ into the formula, 
we obtain the one-loop determinant 
\beq
\frac{\int \D \xi \exp \left( - \frac{1}{2} \int d\tau \, \xi^{\rm T} \Delta \, \xi \right)}{\int \D \xi \exp \left( - \frac{1}{2} \int d\tau \, \xi^{\rm T} \Delta_0 \xi \right)}
= \int d\tau_0 d\phi_0 \, \sqrt{ \det \left( \frac{1}{\pi} M K_+^\dagger K_- \right) } 
= \int d\tau_0 \frac{16i \,  e^{-2i \omega {\rm Im} \, \tau_0} \omega^4}{g^2 (\omega^2-m^2)},
\eeq
where the overall sign has to be determined by taking into account 
how the original path integral is deformed.  
The fact that this determinant is purely imaginary implies 
that there exists an unstable eigenmode of $\Delta$ with a negative eigenvalue. 
We will discuss these issues in the next section. 
Combining the real and complex bion contributions
and using $\lim_{\beta \rightarrow \infty} \frac{1}{\beta} \int d\tau_0 = 1$, 
we obtain
\beq
- \lim_{\beta \rightarrow \infty} \frac{1}{\beta} \frac{Z_1}{Z_0} 
&=& - i( 1 - e^{\pm 2\pi \epsilon i}) \frac{16\omega^4}{g^2(\omega^2-m^2)} 
\exp \left( - \frac{2\omega}{g^2} - 2\epsilon 
\log \frac{\omega+m}{\omega-m} \right), 
\label{eq:one-loop}
\eeq
where we have used Eq.\,\eqref{eq:imaginarytau0} 
to determine the relative sign of the real and complex bion contributions. 
This is the leading order non-perturbative correction to the partition function.
Note that the saddle point method with the Gaussian approximation 
gives the asymptotic form for small ``Planck constant", 
i.e. the inverse of the overall coefficient of the action. 
Therefore, we should regard 
the leading order saddle point result \eqref{eq:one-loop} 
as the asymptotic form of the partition function 
in the weak coupling limit $g \rightarrow 0$
with fixed boson-fermion coupling constant $\lambda \equiv \epsilon m g^2$. 
This result cannot be obtained by conventional methods 
based on the well-separated kink-antikink ansatz (such as the dilute gas approximation)
since the kink-antikink relative distance 
in the saddle point configurations is not necessarily large for fixed $\lambda$
as can be seen from Eqs.\,\eqref{eq:rb_moduli} and \eqref{eq:cb_moduli}.

Although this non-perturbative correction vanishes for $\epsilon=1$ 
as expected from the discussion of the supersymmetry,
it does not reproduce the correct ground state energy 
for the near supersymmetric case \eqref{eq:nearSUSY}.
This is because the weak coupling limit $g \rightarrow 0$ 
with fixed $\lambda$ is equivalent to the large $\epsilon$ limit.
Therefore it does not agree with the near supersymmetric result with $\epsilon \approx 1$.

\subsection{Normalizable quasi zero modes around bion configurations}
In the previous section, 
we have computed the bion contributions by assuming 
that fluctuations around the saddle point configurations are small. 
However, if we take the $g \rightarrow 0$ limit  
without fixing $\lambda=\epsilon m g^2$, 
the determinant \eqref{eq:determinant} vanishes 
and hence low frequency (light) modes emerge in such a limit.  
Since the truncation at the quadratic order is not valid for nearly massless modes, 
their contributions are not fully captured in the one-loop determinant. 
Here, we look for such quasi zero modes
which become nearly massless for small $g$ and $\lambda$. 

First, note that the exact zero modes given in \eqref{eq:Xi_0} 
can be obtained by differentiating the background solution 
with respect to the overall position and phase $(\tau_0, \phi_0)$. 
They can be viewed as superpositions of 
the position and phase modes localized 
on the constituent kink and antikink. 
As can be seen from Eqs.\,\eqref{eq:rb_moduli} and \eqref{eq:cb_moduli}, 
the constituent kink and antikink are 
isolated from each other in the weak coupling limit $\lambda \rightarrow 0$: 
\beq
\tau_+ - \tau_- \, \approx \, \frac{1}{m} \log \frac{2m^2}{\lambda} 
~\rightarrow~ \infty ~~~ ( \lambda \rightarrow 0 ),
\eeq
and hence any superposition of the position and phase modes becomes massless. 
This fact implies that the relative modes corresponding to the relative position and phase 
are nearly massless in the weak coupling regime $\lambda \approx 0$. 
As with the case of the overall zero modes, 
such relative modes can also be represented 
as the derivatives of the background bion solutions
with respect to the relative position $\tau_r \equiv \tau_+ - \tau_-$ 
and phase $\phi_r \equiv \phi_+ - \phi_-$:  
\beq
\xi_{\tau_r} =
\frac{1}{g} 
\ba{cc} 
e_1 & e_2 \\ 
\bar e_1 & \bar e_2 
\ea^{-1}
\ba{c} 
\p_{\tau_r} \varphi_{\rm sol} \\
\p_{\tau_r} \tilde \varphi_{\rm sol} 
\ea,
\hs{10}
\xi_{\phi_r} =
\frac{1}{g} 
\ba{cc} 
e_1 & e_2 \\ 
\bar e_1 & \bar e_2 
\ea^{-1}
\ba{c} 
\p_{\phi_r} \varphi_{\rm sol} \\
\p_{\phi_r} \tilde \varphi_{\rm sol} 
\ea, 
\eeq
where $\varphi_{\rm sol}$ and $\tilde \varphi_{\rm sol}$ depends on 
the positions $\tau_\pm$  and phases and $\phi_\pm$ as
\beq
\varphi_{\rm sol} = \left( e^{\omega(\tau-\tau_+)-i\phi_+} + e^{-\omega(\tau-\tau_-)-i\phi_-} \right)^{-1}, \hs{2}
\tilde \varphi_{\rm sol} = \left( e^{\omega(\tau-\tau_+)+i\phi_+} + e^{-\omega(\tau-\tau_-)+i\phi_-} \right)^{-1}, 
\label{eq:kink_antikink}
\eeq
and $(\tau_\pm, \phi_\pm)$ are set to the values in Eqs.\,\eqref{eq:rb_moduli} and \eqref{eq:cb_moduli}
after differentiating the solution with respect to the relative parameters $(\tau_r, \phi_r)$.
The relative modes $\xi_{\tau_r}$ and $\xi_{\phi_r}$ are 
approximate normalizable eigenmodes 
in the weak coupling regime $\lambda \approx 0$, 
whose eigenvalues are approximately given by
\beq
\Delta \xi_{\tau_r} \approx \lambda \xi_{\tau_r}, \hs{10} 
\Delta \xi_{\phi_r} \approx - \lambda \xi_{\phi_r}.
\eeq
Note that the negative eigenvalue of $\xi_{\phi_r}$ implies that 
the bion solution is unstable under a real variation of $\phi_r$. 

A numerical analysis implies that 
the other fluctuation modes seem to have frequencies higher than $\omega$ 
and hence they are non-normalizable modes with continuous spectra. 
Their contribution to the path integral 
can be taken into account by 
using the one-loop determinant as in the previous subsection. 

On the other hand, as with the case of exact zero modes, 
the contributions of the light normalizable modes
should be calculated by means of the quasi moduli integral, 
i.e. the integral over the quasi moduli parameters $(\tau_r, \phi_r)$:
\beq
Z_{\rm bion} ~\approx~ \int d\tau_0 d\phi_0 \, d\tau_r d \phi_r \, {\det}'' \Delta \exp \left( - V_{\rm eff} \right),
\label{eq:QMI_schematic}
\eeq
where ${\det}'' \Delta$ denotes the functional determinant 
excluding both the exact and the quasi zero modes.  
The effective potential $V_{\rm eff}$ is 
a function of the quasi moduli parameters 
which can be regarded as a reduced action on ``the quasi moduli space", 
i.e. the bottom of valley of the action in the configuration space. 
In the next section, we evaluate the quasi moduli integral \eqref{eq:QMI_schematic}
using an asymptotic form of the effective potential $V_{\rm eff}$.

\section{Bion contributions and quasi moduli integral}
In the previous section, 
we have seen that normalizable modes corresponding to 
the relative position and phase become nearly massless 
in the weak coupling limit $g \rightarrow 0$.
Their contributions may not be fully captured in the one-loop determinant 
since the action is almost flat along the directions corresponding to the nearly massless modes. 
In this section, we apply the Lefschetz thimble formalism to our system
and compute the contributions of the nearly massless modes 
by using complexified quasi moduli integrals.

\subsection{Lefschetz thimble formalism}
Let us first recapitulate the Lefschetz thimble formalism,
which enables us to handle complexified path integrals 
beyond the level of the Gaussian approximation around saddle points.  

In general, a partition function $Z$ defined as a path integral 
can be formally rewritten by deforming the integration contour as
\beq
Z ~=~ \int \D \varphi \, \exp \left(-S[\varphi]\right) ~=~ \sum_{\sigma \in \mathfrak S} n_\sigma Z_\sigma, 
\eeq
where $\mathfrak S$ denotes the set of all saddle points of the action $S$. 
For simplicity, we assume that all the saddle points are non-degenerated, 
i.e. there is no flat direction around each saddle point. 
The contribution from each saddle point $Z_\sigma$ is given by the path integral over 
``the Lefschetz thimble $\mathcal J_\sigma$" 
associated with the saddle point $\sigma$: 
\beq
Z_\sigma \equiv \int_{\mathcal J_\sigma} \D \varphi \, \exp (-S).
\label{eq:Z_sigma}
\eeq
The thimble $\mathcal J_\sigma$ is the set of points in the complexified configuration space 
which can be reached by upward flows from the saddle point $\sigma$: 
\beq
\frac{d \varphi}{dt} = G^{-1} \overline{\frac{\delta S}{\delta \varphi}}, \hs{10}
\lim_{t \rightarrow -\infty} \varphi = \varphi_{\rm sol,\,\sigma}, 
\label{eq:flow}
\eeq
where $t$ is a formal flow parameter and $G$ is the target space metric. 
Note that ${\rm Re} \, S$ is strictly increasing and ${\rm Im} \, S$ is constant along the upward flow 
\beq
\frac{d}{d t} {\rm Re} \, S > 0 , \hs{10} 
\frac{d}{d t} {\rm Im} \, S = 0. 
\eeq
The coefficients $n_\sigma$ indicate
how the original integration cycle $\mathcal C_\R$ of the path integral
is deformed into the union of $\mathcal J_\sigma$: 
\beq
\mathcal C_\R = \sum_{\sigma \in \mathfrak S} n_\sigma \mathcal J_\sigma. 
\eeq
They can also be defined as the intersection numbers between $\mathcal C_{\R}$ 
and ``the dual thimble $\mathcal K_\sigma$" defined as the set of points 
which flows to the saddle point $\sigma$: 
\beq
\frac{d \varphi}{dt} = G^{-1} \overline{\frac{\delta S}{\delta \varphi}}, \hs{10}
\lim_{t \rightarrow \infty} \varphi = \varphi_{\rm sol,\,\sigma}.
\eeq
Since the thimble $\mathcal J_\sigma$ and its dual $\mathcal K_\sigma$ are 
defined in terms of the flow, 
it follows that the real and imaginary parts of the complexified action satisfy
\beq
{\rm Re} \, S |_{\mathcal J_\sigma} \geq {\rm Re} \, S |_{\rm sol, \sigma} \geq {\rm Re} \, S |_{\mathcal K_\sigma}, \hs{10}
{\rm Im} \, S |_{\mathcal J_\sigma} = {\rm Im} \, S |_{\rm sol, \sigma} = {\rm Im} \, S |_{\mathcal K_\sigma}.
\eeq 
These properties imply that $\mathcal J_\sigma$ and $\mathcal K_\sigma$ 
intersect exactly once at the saddle point $\sigma$, 
and $\mathcal J_\sigma$ cannot intersect with $\mathcal K_{\sigma'}~(\sigma' \not = \sigma)$ 
since ${\rm Im} \, S |_{\mathcal J_\sigma} \not = {\rm Im} \, S |_{\mathcal K_{\sigma'}}$ for a generic action. 
Therefore, the intersection pairing of $\mathcal J_\sigma$ and $\mathcal K_{\sigma'}$,
regarded as middle dimensional relative homology cycles, is given by
\beq
\langle \mathcal J_\sigma, \mathcal K_{\sigma'} \rangle = \delta_{\sigma \sigma'}. 
\eeq
Using this pairing, we can calculate the coefficients $n_\sigma$ as the intersection number 
of the original contour $\mathcal C_\R$ and the dual thimble $\mathcal K_\sigma$:
\beq
n_\sigma = \langle \mathcal C_\R , \mathcal K_\sigma \rangle.
\eeq

The perturbative part of the partition function corresponds to 
$Z_0$ defined as the path integral over the thimble $\mathcal J_0$
emanating from the trivial vacuum configuration. 
Non-perturbative contributions are given by the path integral over 
thimbles associated with non-trivial saddle points $\sigma$. 
It is often the case that the partition function 
for a real positive coupling constant $g$ is on the Stokes line, 
i.e. the line on which the thimbles $\mathcal J_\sigma$ 
and the coefficients $n_\sigma$ change discontinuously 
when we vary the coupling constant in the complex $g$ plane. 
If $\mathcal J_0$ jumps on the real axis (${\rm Im} \, g = 0$), 
the perturbative part $Z_0$ has an ambiguity depending on 
how we take the limit ${\rm Im} \, g \rightarrow \pm 0$. 
However, the original partition function $Z$ has no ambiguity 
since it is defined independently of $\mathcal J_\sigma$ and $n_\sigma$. 
Therefore, the ambiguity of $Z_0$ has to be canceled by 
those associated with other non-trivial saddle points. 
In the case of $\C P^1$ quantum mechanics, 
such saddle points correspond to the bion configurations  
and their contributions have an ambiguity 
as can also be seen in the result of the Gaussian approximation \eqref{eq:one-loop}. 
We will see below that the ambiguity of the bion contribution 
originates from the discontinuous change of the intersection number $n_\sigma$ 
associated with the bion saddle points. 

\paragraph{Decomposing degrees of freedom \\}
In the previous section, 
we have calculated the path integral over the thimbles 
$\mathcal J_{\sigma = {\rm rb}}$ and $\mathcal J_{\sigma = {\rm cb}}$ 
by using the leading order Gaussian approximation. 
However, such an approximation is not sufficient for small $g$ and $\lambda$, 
since there emerge nearly massless modes, 
for which the truncation of the fluctuations at the quadratic order is not valid. 
In the following, we evaluate the contributions of the such quasi zero modes 
based on the Lefschetz thimble method. 

To see the thimbles structure in the $\C P^1$ quantum mechanics, 
let us decompose the degrees of freedom as
\beq
\ba{c} \varphi \\ \tilde \varphi \ea = 
\ba{c} \varphi_{k \bar k} \\ \tilde \varphi_{k \bar k} \ea + 
g 
\ba{cc} e_1 & e_2 \\ \bar e_1 & \bar e_2 \ea 
\ba{c} \xi^1 \\ \xi^2 \ea.
\eeq
The background configuration $(\varphi_{k \bar k}, \tilde \varphi_{k \bar k})$ is
a kink-antikink ansatz with the (quasi )moduli parameters $(\tau_0, \phi_0, \tau_r, \phi_r)$
satisfying the equation of motion up to the quasi zero modes: 
\beq
\ba{c}
\displaystyle \phantom{\Bigg|} \frac{\delta S}{\delta \varphi} \phantom{\Bigg|} \\
\displaystyle \phantom{\Bigg|} \frac{\delta S}{\delta \tilde \varphi} \phantom{\Bigg|}
\ea
= a 
\ba{c} 
\displaystyle \phantom{\Bigg|} \frac{\p \varphi_{k \bar k}}{\p \tau_r} \phantom{\Bigg|} \\ 
\displaystyle \phantom{\Bigg|} \frac{\p \tilde \varphi_{k \bar k}}{\p \tau_r} \phantom{\Bigg|} 
\ea 
+ b 
\ba{c} 
\displaystyle \phantom{\Bigg|} \frac{\p \varphi_{k \bar k}}{\p \phi_r} \phantom{\Bigg|} \\ 
\displaystyle \phantom{\Bigg|} \frac{\p \tilde \varphi_{k \bar k}}{\p \phi_r} \phantom{\Bigg|} 
\ea.
\label{eq:quasi_def}
\eeq
Here the coefficients $a$ and $b$ are certain functions of $(\tau_r, \phi_r)$ which vanish at the saddle points. 
Eq.\,\eqref{eq:quasi_def} defines the bottom of the valley of the action
parameterized by the quasi moduli parameters $(\tau_r, \phi_r)$ 
along which the variations of the action in the normal directions vanish.
 
The massive fluctuations $\xi^a$ are 
taken to be orthogonal to the exact and the quasi zero modes. 
Then, Eq.\,\eqref{eq:quasi_def} ensures that the expanded action 
does not have linear term in $\xi$: 
\beq
S[\varphi, \tilde \varphi] ~=~ V_{\rm eff}(\tau_r, \phi_r) + \frac{1}{2} \int d\tau \, \xi^{\rm T} \Delta_{k \bar k} \xi + \mathcal O(g), 
\label{eq:decomp_action}
\eeq
where $\Delta_{k \bar k}$ is the differential operator 
in the background of the kink-antikink configuration. 
For $g \approx 0$, 
the flow for each massive eigenmode in $\xi$ can be approximated as a straight line 
and the integration along such flows gives 
the one-loop determinant of $\Delta_{k \bar k}$.
For the quasi moduli parameters, 
the gradient flow of the action reduces to that of the effective potential $V_{\rm eff}$:
\beq
\frac{d\eta^i}{dt} = \mathcal {G'}^{i \bar j} \, \overline{\frac{\p V_{\rm eff}}{\p \eta^j}},
\eeq
where $\eta^1 = \tau_r,~\eta^2=\phi_r$ and $\mathcal G'_{i \bar j}$ is the metric on the quasi moduli space
\beq
\mathcal G'_{i \bar j} ~\equiv~ \frac{1}{g^2} \int d\tau \frac{1}{(1+|\varphi_{k \bar k}|^2)^2} \frac{\p \varphi_{k \bar k}}{\p \eta^i} \overline{\frac{\p \varphi_{k \bar k}}{\p \eta^j}} . 
\eeq

In the weak coupling limit $\lambda \rightarrow 0$, 
the saddle point configurations can be viewed as well-separated kink-antikink pairs, 
so that the path integral is dominated by the contributions from the well-separated region.
In such a case, $\varphi_{k \bar k}$ and $\tilde \varphi_{k \bar k}$ 
can be well approximated by the simple kink-antikink ansatz \eqref{eq:kink_antikink}. 
Therefore, the flows around the saddle points can be determined 
by using the following asymptotic form of the effective potential obtained by substituting 
the ansatz \eqref{eq:kink_antikink} into the action:
\beq
V_{\rm eff} \approx \frac{2m}{g^2} - \frac{4m}{g^2} e^{-m \tau_r} \cos \phi_r + 2 \epsilon m \tau_r. 
\label{eq:Veff}
\eeq
Therefore, the bion contributions for small $g$ and $\lambda$ can be 
obtained by applying the Lefschetz thimble method to the quasi moduli integral 
\beq
\frac{Z_1}{Z_0} &\approx& 
\int d\tau_0 d\phi_0 \, d\tau_r d \phi_r 
\sqrt{\det \left( \frac{\mathcal G}{2\pi} \right) \det \left( \frac{\mathcal G'}{2\pi} \right) \frac{ \, \det \, \Delta_0}{{\det}'' \Delta_{k \bar k} }} 
\exp \left( - V_{\rm eff} \right).
\eeq
Since the spectrum of $\Delta_{k \bar k}$ for a well-separated kink-antikink configuration 
can be approximated as two copies of that of a single kink, 
we find that
\beq
\sqrt{ \frac{\det \, \Delta_0 \, }{{\det}'' \Delta_{k \bar k}} } 
~=~ \frac{\int \D \xi \exp \left(- \frac{1}{2} \int d\tau \, \xi^{\rm T} \Delta_{k \bar k} \xi \right)}{\int \D \xi \exp \left(- \frac{1}{2} \int d\tau \, \xi^{\rm T} \Delta_{\, 0 \,} \xi \right) }
~\approx~ \frac{\det \Delta_0}{{\det}' \Delta_{\rm k}}. 
\eeq
The one-loop determinant in the single kink background ${\det}' \Delta_{\rm k}$ 
can be obtained by using the formula \eqref{eq:determinant} as
\beq
\frac{\det \Delta_0}{{\det}' \Delta_{\rm k}} ~=~ \frac{1}{\det \mathcal G_k} \left( \frac{4m^2}{g^2} \right)^2,
\eeq
where $\mathcal G_k$ is the metric on the single kink moduli space. 
The overall and the relative moduli space metrics also reduce 
to the two copies of the single kink metric
\beq
\sqrt{\det \mathcal G \det \mathcal G'} ~ \approx ~ \det \mathcal G_k. 
\eeq
In summary, the leading order bion contribution 
to the ground state energy for small $g$ and $\lambda$
is given by the following quasi moduli integral
\beq
- \lim_{\beta \rightarrow \infty} \frac{1}{\beta} \frac{Z_1}{Z_0} &\approx& - \frac{8m^4}{\pi g^4} \int d\tau_r d \phi_r \, \exp \left( - V_{\rm eff} \right). 
\label{eq:correction_to_energy}
\eeq
In the following, we apply the Lefschetz thimble method 
to evaluate the quasi moduli integral with the asymptotic potential \eqref{eq:Veff}.

\subsection{Quasi-moduli Integral in sine-Gordon quantum mechanics}
\label{sec:QM}
\setcounter{paragraph}{0}

Before calculating the quasi moduli integral in the $\C P^1$ model, 
let us briefly discuss the case of the sine-Gordon model, 
which can be obtained by restricting the $\C P^1$ mechanics to the zero angular momentum ($l=0$) sector:
\beq
L = \frac{1}{4g^2} \left( \dot \theta^2 - m^2 \sin^2 \theta \right) + \epsilon m \cos \theta, \hs{10}
\left( \varphi = \tan \frac{\theta}{2} e^{i \phi} \right).
\label{eq:SG_acton}
\eeq
This model is not only simpler, 
but also serves as a useful building block for the $\C P^1$ model. 
Since the bion solutions in the $\C P^1$ model are in the zero angular momentum sector,
the sine-Gordon action \eqref{eq:SG_acton} also 
has the corresponding real and complex bion solutions \cite{Behtash:2015zha, Behtash:2015loa}. 
As shown in \cite{Bogomolny:1980ur, ZinnJustin:1981dx, Dunne:2012ae, Misumi:2014jua, Misumi:2015dua}, 
the bion contribution with periodic potentials can be expressed 
in terms of the following quasi zero mode integral 
with respect to the relative separation $\tau$ 
(the subscript of $\tau_r$ will be omitted in the following)
\begin{align}
[{\mathcal I}\bar{\mathcal I}] \,=\, \int_{\mathcal C_{\R}}
d\tau \, e^{-V_{\rm SG}(\tau)} \hs{10}
V_{\rm SG}(\tau) \,\equiv\, - \frac{4m}{g^2} e^{-m \tau} + 2 \epsilon m \tau \,,
\label{oriInt}
\end{align}
where $[{\mathcal I}\bar{\mathcal I}]$ denotes the single bion contribution 
excluding the classical part $e^{-S_{bion}} = e^{-\frac{2m}{g^2}}$.  
The original integration cycle is the real axis 
since the original path integral is over real configurations. 
Note that the divergence of the integrand (\ref{oriInt}) 
in the $\tau \rightarrow -\infty$ limit is an artifact of the approximation, 
since the effective potential is valid only for large $\tau$. 
In the Bogomolny--Zinn-Justin (BZJ) prescription, 
with which we can extract physical information form \eqref{oriInt}, 
we first regard $g^2$ as a negative number and perform the integral.
In the end of the calculation, 
we analytically continue $g^2$ back to a positive number.   
Then we end up with a bion contribution $[{\mathcal I}\bar{\mathcal I}]$ 
with an imaginary ambiguity depending on 
how the result is analytically continued to real positive $g^2$. 
It is known that this ambiguity is cancelled out by another 
imaginary ambiguity emerging from the Borel re-summation of the perturbative series.
This is the well-known first step to the resurgence theory.
In the reference \cite{Behtash:2015loa}, 
it has been indicated that the complexified quasi moduli integral 
is a more rigorous way to treat the quasi moduli integral 
which can replace the BZJ prescription.
We here extend this idea and clarify the BZJ prescription 
in terms of the complexification method.

As we have done in Sec.\,\ref{subsec:complexbion}, 
we first introduce a small complex factor as a regulator into the coupling as
\begin{equation}
g^2 \, \to \, g^2 e^{i\theta}\,.
\end{equation}
In the end of calculation we take $\theta\to\pm0$ limits.
Now, we complexify the quasi moduli $\tau$
\begin{equation}
\tau\in{\mathbb R}\,\,\,\to\,\,\,\tau = \tau_{R}\,+\,i \tau_{I} \in{\mathbb C}\,,
\end{equation}
with the real part $\tau_{R}$ and the imaginary part $\tau_{I}$.
Let $\mathfrak S$ be the set of saddle points in the complex $\tau$-plane. 
Then the original integration cycle can be decomposed as
\beq
{\mathcal C}_{\mathbb R} = \sum_{\sigma \in \mathfrak S} n_{\sigma} {\mathcal J}_{\sigma},
\label{eq:contour_decomposed}
\eeq
where ${\mathcal J}_{\sigma}$ is the thimble associated with the saddle point $\tau_{\sigma}$, 
i.e. the flow line starting from the the saddle point $\tau_{\sigma}$: 
\begin{equation}
\frac{d \tau}{d t} = \frac{1}{2m} \overline{\frac{\p V_{\rm SG}}{\p \tau}} \,, \hs{10}
\lim_{t \rightarrow -\infty} \tau = \tau_{\sigma}, 
\end{equation}
where we have rescaled the flow time parameter $t$ for notational convenience. 
Using the decomposition of the integration contour \eqref{eq:contour_decomposed}, 
we can rewrite the original integral in (\ref{oriInt}) as
\begin{align}
[{\mathcal I}\bar{\mathcal I}] \,
=\,\sum_{\sigma}n_{\sigma}Z_{\sigma}, \hs{10}
Z_{\sigma}\,=\, \int_{{\mathcal J}_{\sigma}} 
d\tau \, e^{-V_{\rm SG}(\tau)}\,,
\end{align}
The coefficients $n_{\sigma}$ is the intersection number
between the original integration cycle ${\mathcal C}_{\mathbb R}$ 
and the dual cycle ${\mathcal K}_{\sigma}$ (dual thimble)
\begin{equation}
n_{\sigma}\,=\, \langle {\mathcal C}_{\mathbb R},\,{\mathcal K}_{\sigma} \rangle
\end{equation}

\paragraph{Thimbles and Dual Thimbles \\}
The effective kink-antikink potential deformed by $\theta$ is written as
\beq
V_{\rm SG}(\tau) = - \frac{4m}{g^2} e^{-m \tau - i\theta} + 2 \epsilon m \tau.
\eeq
What we need to do is just to find the saddle points of the potential, 
and the thimbles and the dual thimbles associated with them.

The saddle points of the potential $V_{\rm SG}$ is labeled by an integer $\sigma \in \Z$:
\begin{align}
\tau \ = \  \tau_\sigma \ \equiv \ \frac{1}{m} \left[ \log \frac{2m}{\epsilon g^2} + (2\sigma-1) \pi i - i \theta \right] \,. 
\label{eq:criticalPoint}
\end{align}
The gradient flow equation is given by
\beq
\frac{d \tau}{d t} ~=~ \frac{1}{2m} \overline{\frac{\p V_{\rm SG}}{\p \tau}} 
~=~ - \frac{2 m}{g^2} e^{m \bar \tau + i\theta} + \epsilon \,.
\eeq
This equation can be solved as 
\beq
\tau(t) = \frac{1}{m} \log \left[ \frac{2m}{\epsilon g^2} 
\frac{\sin(a - b e^{-\epsilon m t} - \theta)}{b e^{-\epsilon m t}} \right] 
- \frac{i}{m} (a - b e^{-\epsilon m t} ),
\label{eq:tau_sol}
\eeq
where $a$ and $b$ are real integration constants. 
The dual thimbles can be defined as 
flows reaching the saddle points at $t\to \infty$. 
The above solution of the flow equation approach 
the saddle point $\tau_{\sigma}$ in Eq.\,(\ref{eq:criticalPoint}),
if the integration constant $a$ is given by
\beq
a_\sigma = -(2\sigma-1) \pi + \theta, \quad \sigma \in \Z.
\eeq
Eliminating $t$ from the solution (\ref{eq:tau_sol}) and its complex conjugate, 
we find that the real part $\tau_R = {\rm Re} \, \tau$ 
and the imaginary part $\tau_I = {\rm Im} \, \tau$ 
are related in the dual thimble $\mathcal K_\sigma$ as 
\beq
m \tau_R = \log \left[ \frac{2m}{\epsilon g^2} 
\frac{\sin(m \tau_I + a_\sigma)}{m \tau_I+a_\sigma} \right], \hs{10} 
(-a_\sigma - \pi \leq m \tau_I \leq -a_\sigma + \pi).
\label{dualthimble1d}
\eeq

On the other hand, thimbles are defined by reaching the 
saddle point at $t \to - \infty$. 
Such solutions can be obtained from 
the general solution \eqref{eq:tau_sol} by redefining the integration constants as
\beq
a = a_\sigma + \delta, \hs{10} b = - \delta e^{-\epsilon m t_0},
\eeq
and taking the limit $\delta \rightarrow 0$:
\begin{equation}
\tau(t) = \frac{1}{m} \log \frac{2m}{\epsilon g^2} \left(1+ e^{\epsilon m (t-t_0)} \right) - \frac{i}{m} a_\sigma.
\label{eq:thimble1d}
\end{equation}
Therefore, the thimble $\mathcal J_\sigma$ is the straight line with fixed imaginary part
\beq
m \tau_I ~=~ - a_\sigma ~=~ (2\sigma-1) \pi - \theta.
\label{eq:thimble1d}
\eeq

\paragraph{Integral along Lefschetz Thimbles and BZJ prescription \\}

\begin{figure}[htbp]
\begin{center}
\includegraphics[width=0.48\textwidth]{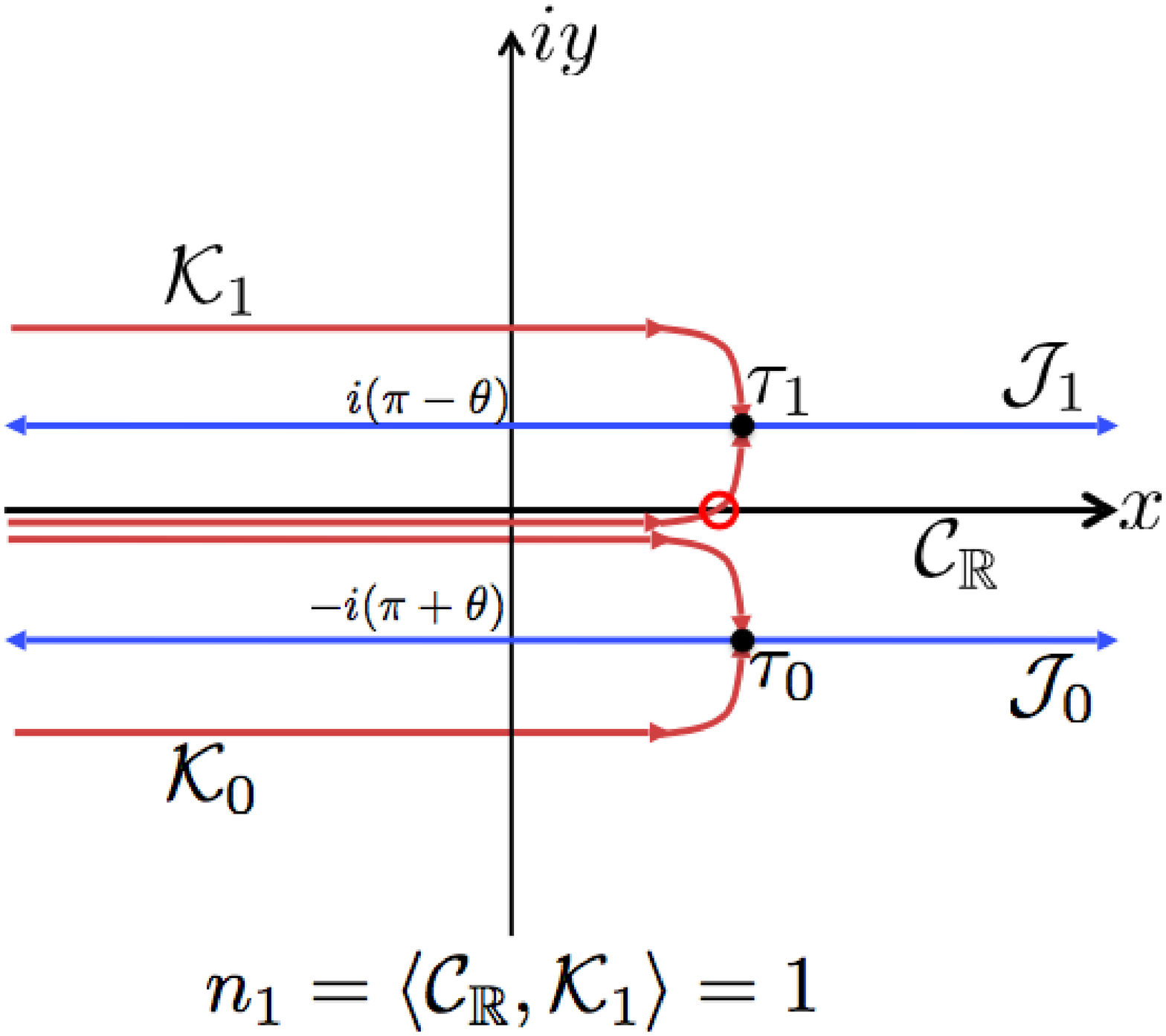}
\includegraphics[width=0.48\textwidth]{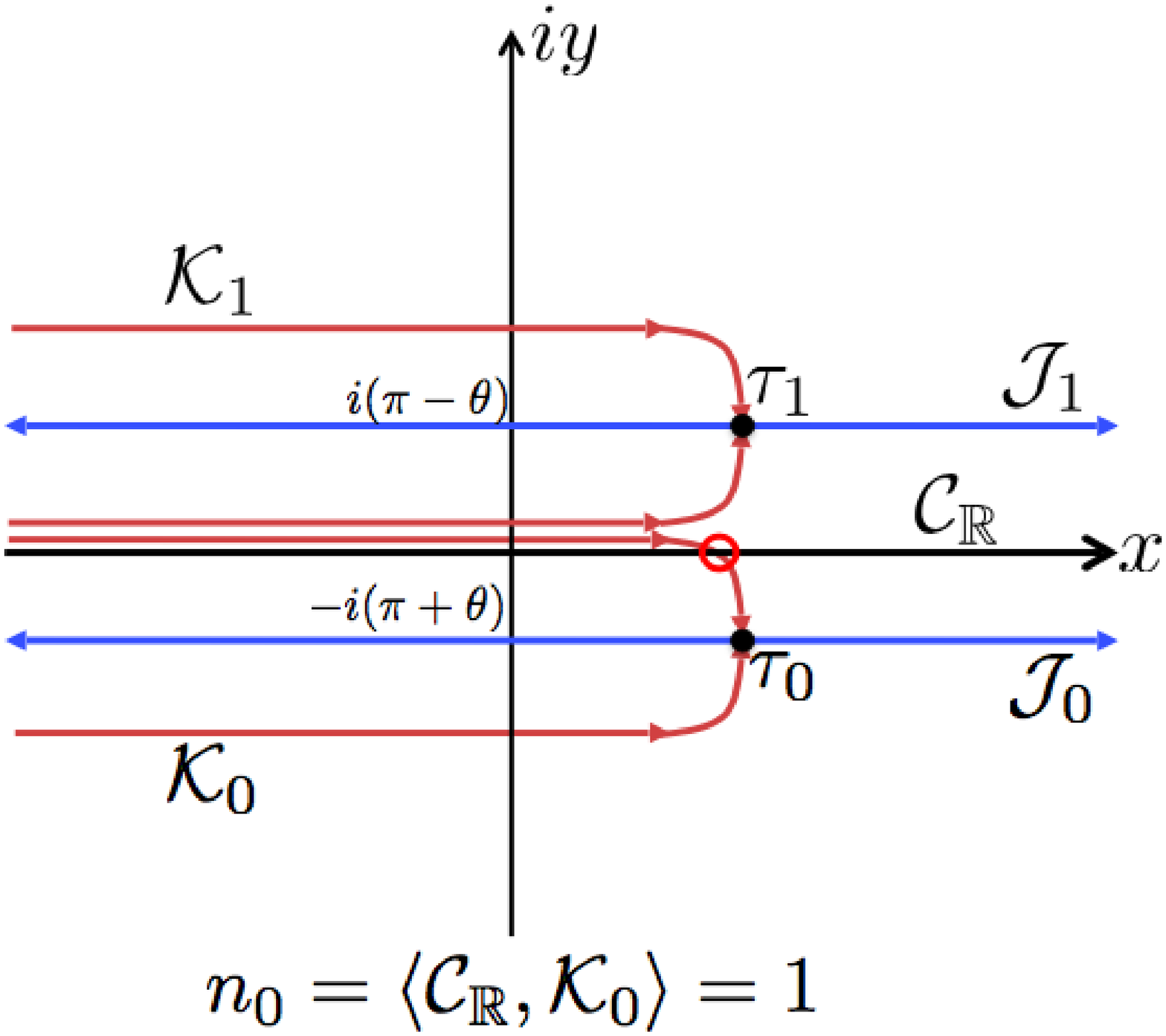}
\end{center}
\caption{Complex integration cycles for bion amplitude in quantum mechanics.
The regularization parameter is $\theta = +0$(left) and $\theta=-0$(right).}
\label{fig:thimble1}
\end{figure}

In Fig.\,\ref{fig:thimble1}, the two critical points $\tau_{\sigma}$, 
the associated thimbles ${\mathcal J}_{\sigma}$ 
and the dual cycles ${\mathcal K}_{\sigma}$ are depicted for $\sigma=0$ and $\sigma=1$.
For $\theta=0$, the intersection numbers are ill-defined 
since the dual thimbles never cross  
the original integration path ${\mathcal C}_{\mathbb R}$ although 
they are asymptotically tangent to ${\mathcal C}_{\mathbb R}$. 
This is one of the reasons why we need to regulate the potential 
by the imaginary part of the coupling constant.

As shown in Fig.\,\ref{fig:thimble1}, 
the intersection number jumps when $\theta$ changes its sign:
\beq
(n_0, n_1) = \left\{ \begin{array}{cc} (0,1) & \mbox{for $\theta = +0$} \\ (1,0) & \mbox{for $\theta=-0$} \end{array} \right.,
\eeq
and hence
\beq
[{\mathcal I}\bar{\mathcal I}] 
~=~ \sum_{\sigma} n_\sigma Z_\sigma 
~=~ \left\{ \begin{array}{cc} 
Z_{\sigma=1} & \mbox{for $\theta = +0$} \\ 
Z_{\sigma=0} & \mbox{for $\theta=-0$} 
\end{array} \right..
\eeq
Integrating along each thimble $\mathcal J_\sigma$, 
we find that
\beq
Z_\sigma ~=~ \int_{\mathcal J_\sigma} d\tau \, \exp \left(- V_{\rm SG} \right) 
~=~ \frac{1}{m} e^{- 2\pi i \epsilon (2 \sigma - 1)} \left( \frac{g^2}{4m} e^{i \theta} \right)^{2\epsilon} \Gamma(2\epsilon).
\eeq
This agrees with the result from the BZJ description. 
In this calculation of the complex integration, 
the region where the integrand is divergent 
is avoided by moving the integration contour $[-\infty,\infty]$ 
to either of the Lefschetz thimbles 
$[-\infty\pm i\pi, \infty\pm i\pi]$.
This is how one extracts a finite result from the ill-defined integral in the BZJ prescription.
Now, it is clear that the ambiguity comes from the sign 
of regularization parameter $\theta$ in $g^2 \to g^2 e^{i\theta}$.

To sum up, unambiguous definition of moduli integral is obtained 
by making the coupling constant complex 
and using the Lefschetz thimble approach. 
We can regard this method as a more rigorous definition 
of the BZJ prescription.


\subsection{Bion contributions in ${\mathbb C}P^{1}$ model}
\label{sec:CP}
\setcounter{paragraph}{0}
\paragraph{Quasi-moduli Integral \\}

As discussed above and in \cite{Misumi:2014jua, Misumi:2015dua}, 
the bion contribution in ${\mathbb C}P^{1}$ quantum mechanics can be expressed 
by the following quasi zero mode integral with 
respect to the separation $\tau$ and the relative phase $\phi$,
\beq
[{\mathcal I}\bar{\mathcal I}] ~=~ 
\int_{-\pi}^{\pi} d\phi \int_{-\infty}^{\infty} d\tau \, e^{-V(\tau,\phi)}, \hs{10}
V(\tau,\phi) \ =\ - \frac{4m}{g^2} \cos\phi \, e^{-m \tau} \,+\, 2 \epsilon m \tau \,.
\label{IbarIInt}
\eeq

\paragraph{Thimbles and Dual Thimbles \\}

\begin{figure}[t]
\begin{center}
\begin{minipage}[c]{0.45\hsize}
\includegraphics[width=60mm]{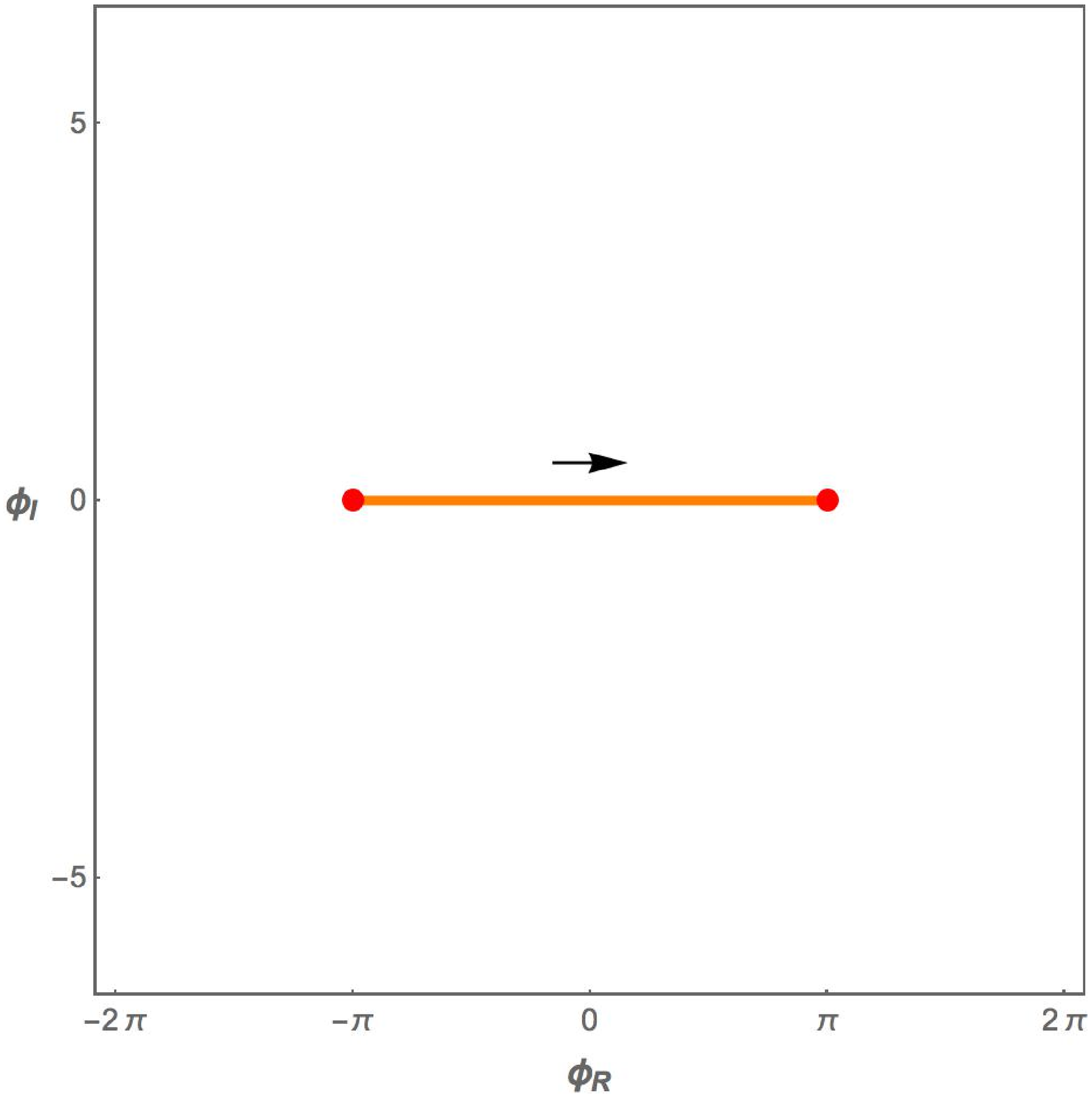} \\
(a) original contour
\end{minipage}
\begin{minipage}[c]{0.45\hsize}
\includegraphics[width=60mm]{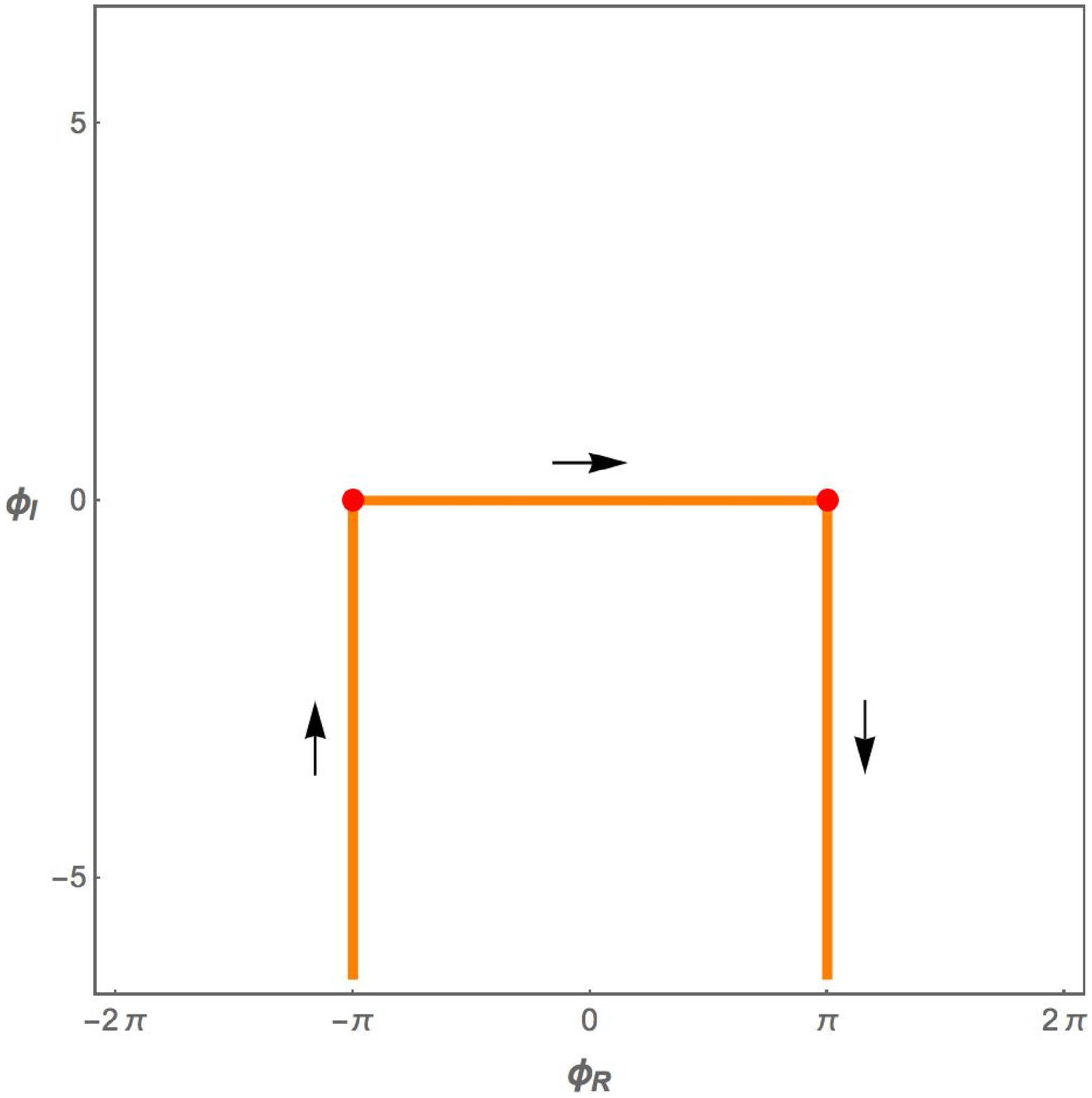} \\
(b) extended contour
\end{minipage}
\caption{Integration contours on the complex $\phi$-plane.}
\label{fig:contours}
\end{center}
\end{figure}

In the integral in question, 
we first consider the following deformation of 
the integration contour for the phase $\phi$ \cite{Witten:2010cx}.
As shown in Fig.\,\ref{fig:contours}, 
we extend the integration path on the $\phi$-plane 
to make it an integration cycle without boundary 
in the complex $\phi$ plane. 
Since the contributions from the paths attached to 
$\phi= \pm \pi$ (the boundaries of the original path) 
cancel with each other, 
the extended contour gives the same value of the bion contribution $[{\mathcal I}\bar{\mathcal I}]$.

\begin{figure}[h]
\begin{center}
\begin{minipage}[c]{0.45\hsize}
\centering
\includegraphics[width=90mm]{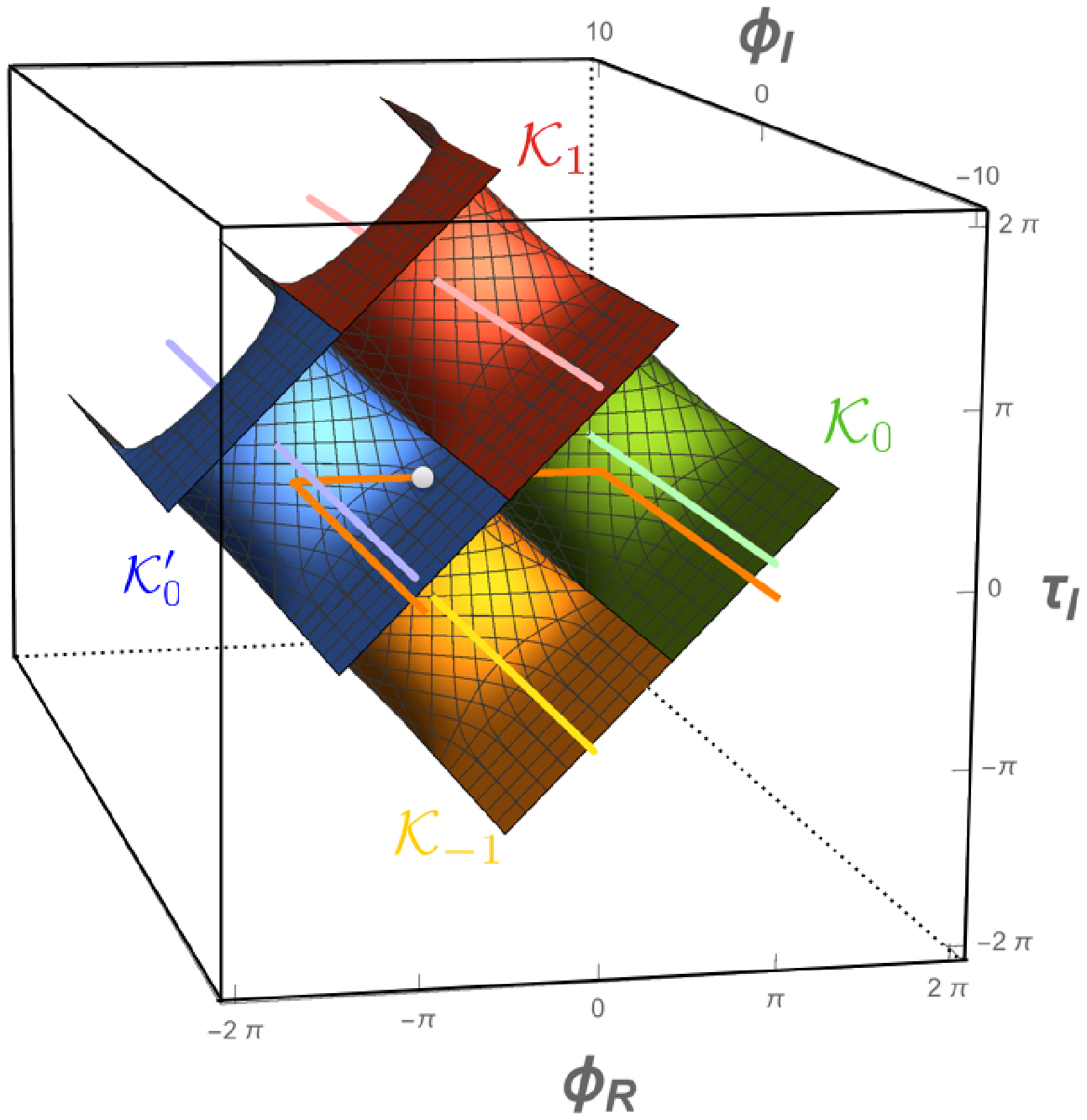} \\
(a) $\theta = -0$
\end{minipage}
\hs{5}
\begin{minipage}[c]{0.45\hsize}
\centering
\includegraphics[width=95mm]{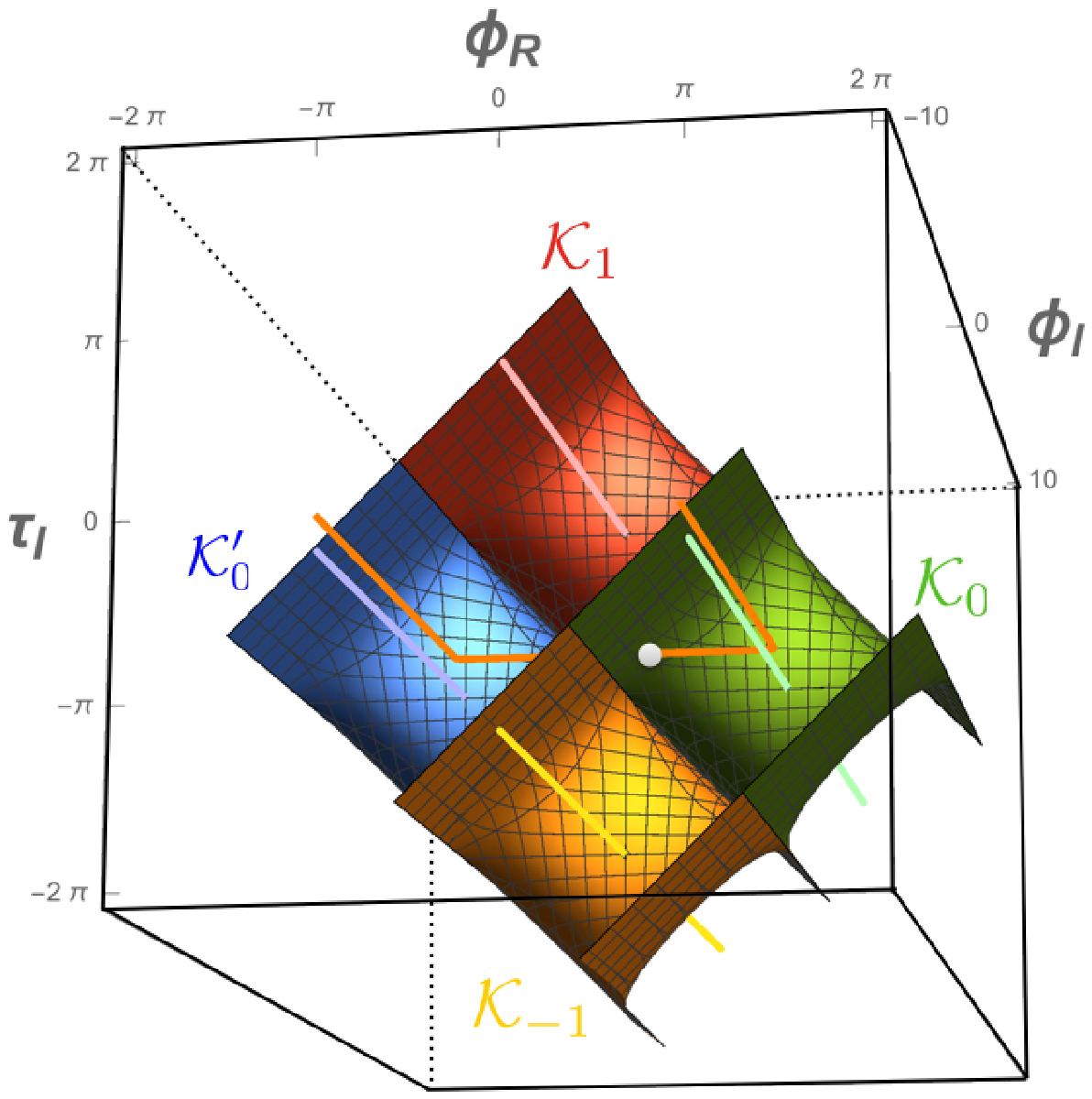} \\
(b) $\theta = +0$
\end{minipage} 
\hs{5}
\caption{Integration contour, Lefschetz thimbles and dual thimbles 
for (a) $\theta = -0$ (seen from upper left)
and (b) $\theta = +0$ (seen from lower right). 
The orange lines are the original extended integration contours,  
while four colored (red, blue, yellow and green) lines and surfaces indicate the thimbles and their duals, respectively.
Note that since the integration contour and Lefschetz thimbles are 
direct products of the $\tau_R$ direction and lines in $(\phi_R,\phi_I,\tau_I)$, 
their projected images are lines in the three-dimensional space. }
\label{fig:dual_thimble}
\end{center}
\end{figure}

We next introduce a small phase 
for the coupling constant as $g^2 \,\to\, g^2 e^{i\theta}$, and
complexify both $\tau$ and $\phi$ as
\begin{equation}
\tau = \tau_{R}+i\tau_{I} \in {\mathbb C}, \hs{10} \phi=\phi_{R}+i\phi_{I} \in {\mathbb C}.
\end{equation}
The saddle points are given by the equations,
\begin{align}
&{\partial V \over{\partial\tau}}\,=\, 
{4m^2 \cos\phi\over{g^2}}e^{-m\tau-i\theta}+2\epsilon m \,=\,0\\
&{\partial V \over{\partial\phi}}\,=\, 
{4m \sin\phi\over{g^2}}e^{-m\tau-i\theta} \,=\,0\,.
\end{align}
The solutions are labeled by a integer $\sigma \in \Z$
\begin{align}
\tau_{\sigma} ~=~ \frac{1}{m} \log \frac{2m}{\epsilon g^2} + \frac{i}{m} (\sigma \pi - \theta), \hs{10} 
\phi_{\sigma} ~=~ - (\sigma - 1)\pi ~~({\rm mod} \, 2\pi) \,.
\label{eq:critical}
\end{align}
The physical interpretation of these saddle point will discussed below. 
To find the thimble $\mathcal J_\sigma$ and its dual $\mathcal K_\sigma$, 
let us redefine the coordinates as
\beq
\tau_+ = \tau + \frac{i}{m} \phi, \hs{10} \tau_- = \tau - \frac{i}{m} \phi.
\eeq
Then we can rewrite the effective potential $V$ 
into two copies of the effective kink-antikink potential 
in the sine-Gordon model \eqref{oriInt}:
\beq
V = \frac{V_{\rm SG}(\tau_+) + V_{\rm SG}(\tau_-)}{2}, 
\eeq
Thus, the solution of the flow equation 
\beq
\frac{d \tau}{dt} = \frac{1}{2m} \overline{\frac{\p V}{\p \tau}}, \hs{10}
\frac{d \phi}{dt} = \frac{m}{2} \overline{\frac{\p V}{\p \phi}}.
\eeq
can be easily obtained as 
\beq
\tau_i &=& \frac{1}{m} \log \left[ \frac{2m}{\epsilon g^2} 
\frac{\sin(a_i - b_i e^{- \epsilon m t} - \theta)}
{b_i e^{-\epsilon m t}} \right] - \frac{i}{m} (a_i - b_i e^{-\epsilon m t} )\,,
\eeq
where $i=(+,-)$ and $a_{i}$ and $b_i$ are integration constants.
The flows on the dual thimbles associated with the critical points \eqref{eq:critical}
can be obtained by setting the integration constants as
\beq
a_+ \ = \ a_{+\sigma} \ \equiv \ - \pi + \theta, \hs{10} 
a_- \ = \ a_{-\sigma} \ \equiv \ - (2\sigma - 1) \pi + \theta\,,
\eeq
Eliminating the parameters $b_1$ and $b_2$, 
we obtain the following equations for the dual thimbles 
\beq
m \tau_R - \phi_I  &=& 
\log \left[ \frac{2m}{\epsilon g^2} 
\frac{\sin(m \tau_I + \phi_R + a_{+\sigma})}{m\tau_I + \phi_R + a_{+\sigma}} \right], \hs{5}
- \pi \leq m \tau_I + \phi_R + a_{+\sigma} \leq \pi, \\
m \tau_R + \phi_I &=& 
\log \left[ \frac{2m}{\epsilon g^2} 
\frac{\sin(m\tau_I - \phi_R +a_{-\sigma})}{m\tau_I - \phi_R + a_{-\sigma}} \right], \hs{5}
- \pi \leq m \tau_I - \phi_R +a_{-\sigma} \leq \pi.
\eeq
The thimbles can also be determined 
by using the result for the sine-Gordon case \eqref{eq:thimble1d} as 
\beq
{\rm Im} \, \tau_+ \ = \ - a_{+\sigma} \ = \ \pi - \theta, \hs{10}
{\rm Im} \, \tau_- \ = \ - a_{-\sigma} = (2\sigma-1) \pi - \theta.
\eeq
Therefore, the thimbles are the planes specified by
\beq
m \tau_I = \sigma \pi - \theta, \hs{10} \phi_R = - (\sigma-1) \pi. 
\eeq
The thimbles and dual thimbles projected onto 
$(\phi_R,\phi_I,\tau_I)$ are shown in Fig.\,\ref{fig:dual_thimble}. 
They have the same structure as in the sine-Gordon case on the two dimensional slices 
$\tau_I + \phi_R = - a_{+\sigma}$ and $\tau_I - \phi_R = - a_{-\sigma}$. 

As with case of the sine-Gordon quantum mechanics, 
the dual thimbles $\mathcal K_1$ and $\mathcal K_0$ intersect 
the original integration contour for $\theta=+0$ 
while the dual thimbles $\mathcal K_{-1}$ and $\mathcal K_0'$ 
cross the original contour for $\theta=-0$. 
Note that $\mathcal K_0$ and $\mathcal K_0'$ are 
the identical thimble related by the shift $\phi_R \rightarrow \phi_R + 2\pi$. 
\begin{figure}[h]
\begin{center}
\begin{minipage}[c]{0.45\hsize}
\includegraphics[height=60mm]{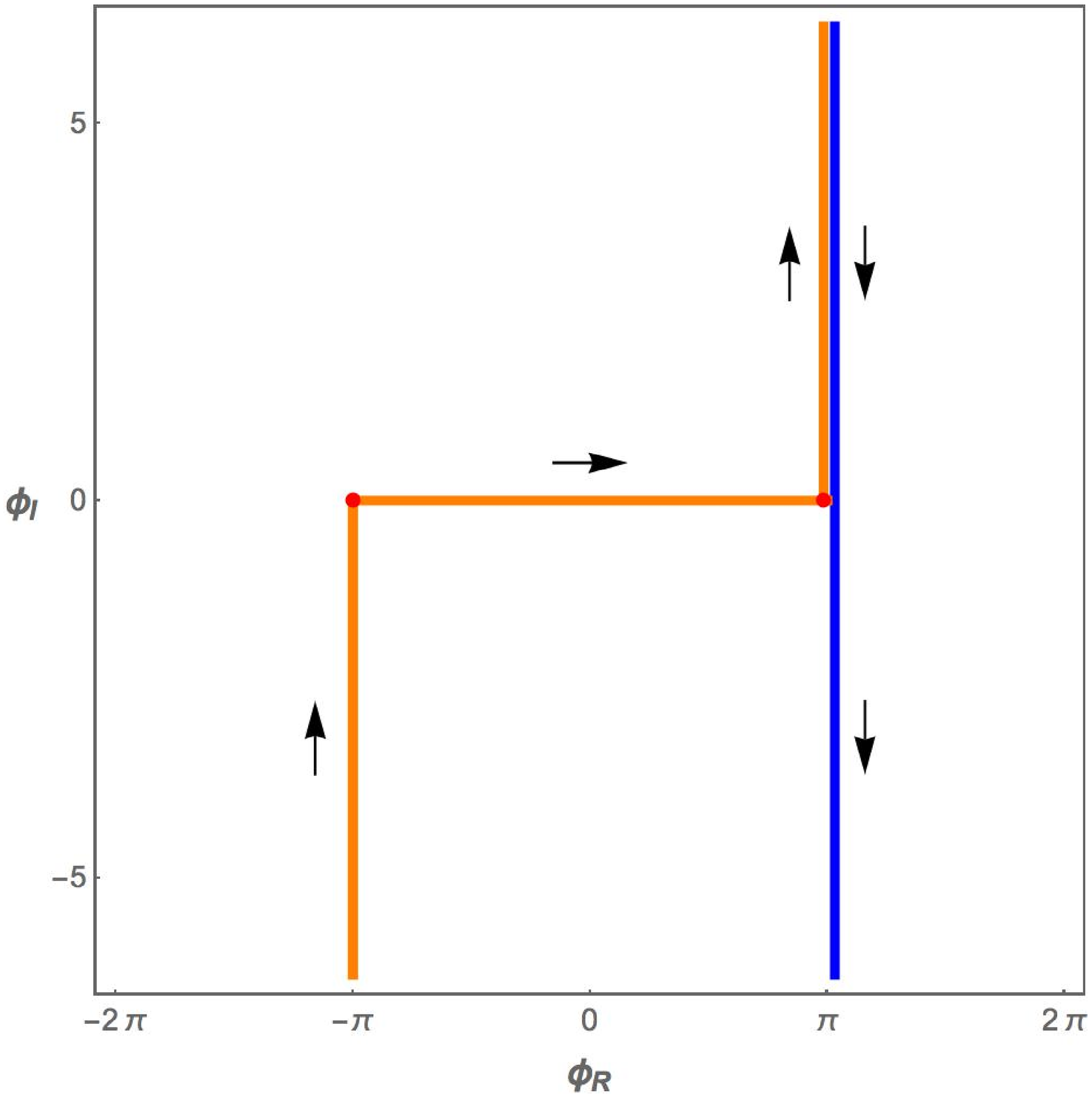} \\
(a) integration contour for $\phi$ 
\end{minipage}
\hs{10}
\begin{minipage}[c]{0.45\hsize}
\includegraphics[height=60mm]{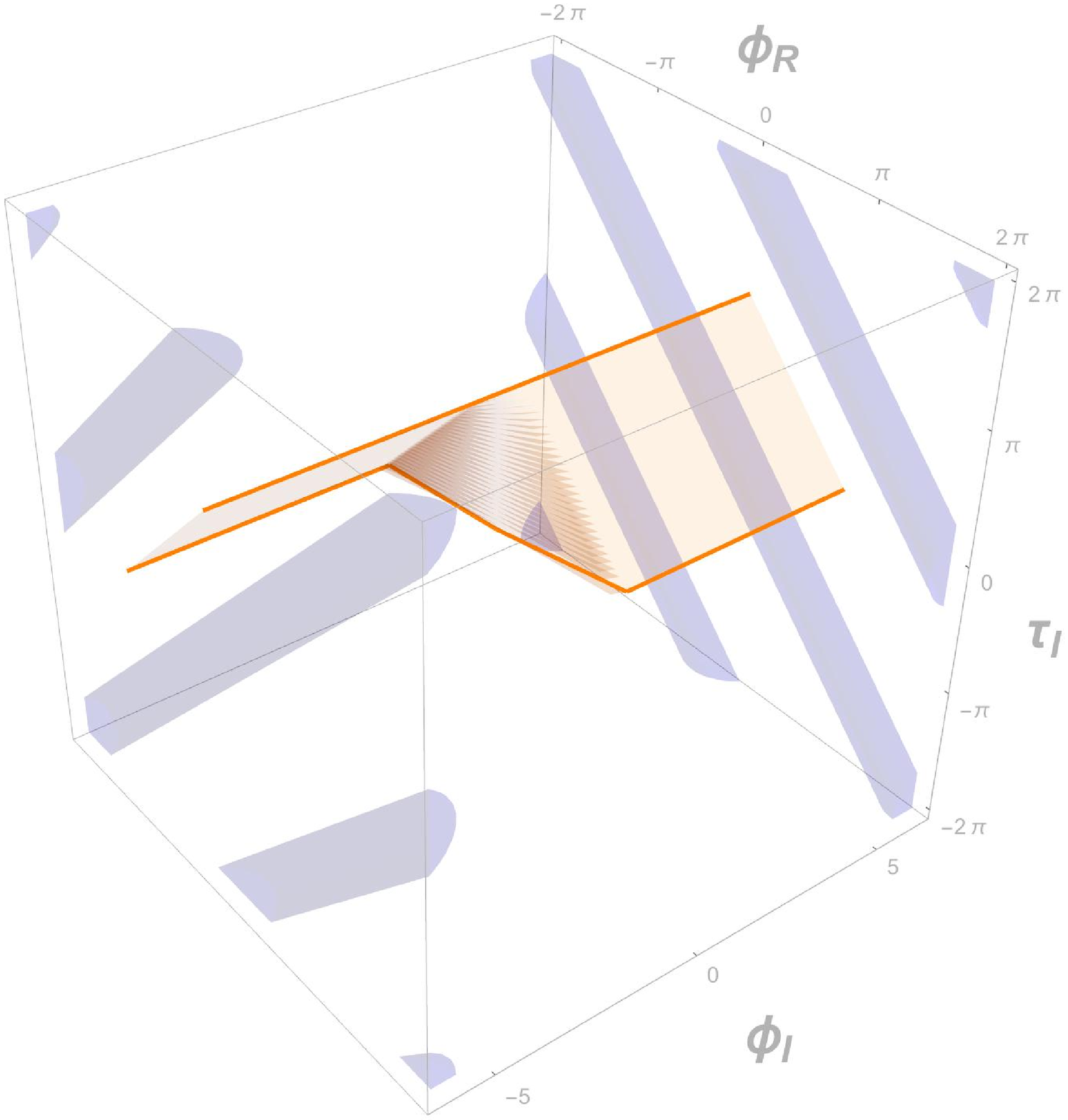} \\
(b) deformation of contour
\end{minipage}
\caption{Deformation of integration contour. 
(a) The integration contour can be decomposed into two paths 
(orange and blue). 
One of them corresponds to the thimble with $n=0$ and 
the other can be continuously deformed into the thimble with $n=1$. 
The shaded regions in the right figure corresponds to the 
region where ${\rm Re} \, V < T \ll {\rm Re} \, V_{\rm critical}$ with some real number $T$. }
\label{fig:contour_deformation}
\end{center}
\end{figure}
By taking into account how the original integration contour 
is decomposed into the thimbles (see Fig.\,\ref{fig:contour_deformation}), 
the sign of the intersection numbers can be determined as
\beq
(n_{-1} \, , \, n_{0} \, , \, n_{1}) =
\left\{ 
\begin{array}{cc}
(\, -1 \, , \, 1 \, , \, 0 \,) ~&~ \mbox{for $\theta=-0$} \\
(\, 0 \, , \, -1\, , \, 1 \,) ~&~ \mbox{for $\theta=+0$}
\end{array} 
\right..
\eeq
Therefore, the bion contribution has the ambiguity depending on the sign of $\theta$
\beq
[{\mathcal I}\bar{\mathcal I}] ~=~ 
\left\{ 
\begin{array}{cc}
Z_{\sigma=0} - Z_{\sigma=-1} \, ~&~ \mbox{for $\theta=-0$} \\
Z_{\sigma=1} - Z_{\sigma=0} ~&~ \mbox{for $\theta=+0$}
\end{array} 
\right..
\eeq

\paragraph{Integral along Lefschetz Thimbles \\}
Now let us evaluate the integral over the thimbles.
Changing the coordinates as
\beq
\tau \rightarrow \tau' = \tau - \tau_\sigma \hs{10} 
\phi \rightarrow \phi' = \phi - \phi_\sigma. 
\eeq
we find that the potential becomes
\beq
V = 2 \epsilon \left( m \tau' + e^{-m\tau'} \cos \phi' + \log \frac{2m}{\epsilon g^2} + \sigma \pi i - i \theta \right).
\eeq
The thimble $\mathcal J_\sigma$ corresponds to the two dimensional plane $\tau' \in \R,~\phi' \in i \R$. 
We can check that the potential satisfies
\beq
{\rm Re} \, V \geq 2 \epsilon \left( 1 + \log \frac{2m}{\epsilon g^2} \right), \hs{10} 
{\rm Im} \, V = (\sigma \pi - \theta) \epsilon = const.,
\eeq
for $\tau' \in \R,~\phi' \in i \R$. 
Integrating over the thimbles, we obtain
\beq
Z_\sigma \ = \ \int_{\R} d \tau' \int_{i \R} d \phi' \, e^{-V} \ = \ \frac{i}{2m} \left( \frac{g^2 e^{i\theta}}{2m} \right)^{2\epsilon} e^{-2 \pi i \epsilon \sigma} \, \Gamma \left( \epsilon \right)^2.
\eeq
Therefore, the bion contribution is given by
\beq
[{\mathcal I}\bar{\mathcal I}] ~=~ 
\frac{1}{m} \left( \frac{g^2 e^{i\theta}}{2m} \right)^{2\epsilon} \sin \epsilon \pi \, \Gamma \left( \epsilon \right)^2 \times
\left\{ 
\begin{array}{ll}
e^{\pi i \epsilon} ~&~ \mbox{for $\theta=-0$} \\
e^{-\pi i \epsilon} ~&~ \mbox{for $\theta=+0$}
\end{array} 
\right..
\eeq
This result is consistent with the one obtained by 
applying the Bogomolny--Zinn-Justin prescription for 
the divergent region $\tau \rightarrow - \infty$, $|\phi| \le \pi/2$ \cite{Misumi:2015dua}. 
In this calculation of the complex integral, 
the region where the integrand is divergent 
is avoided by deforming the integration contour 
as shown in Fig.\,\ref{fig:contour_deformation}.
This is how one extracts a finite result from the ill-defined integral in the BZJ prescription.
Thus, based on the Lefschetz thimble decomposition of 
the quasi moduli integral together with the complexification of the coupling, 
we obtain an unambiguous definition of the ill-defined moduli integral. 

From Eq.\,\eqref{eq:correction_to_energy} and 
the result of the complexified quasi moduli integral, 
we obtain the following non-perturbative correction to the ground state energy:  
\beq
- \lim_{\beta \rightarrow \infty} \frac{1}{\beta} \frac{Z_1}{Z_0} &\approx& 
- \frac{8m^4}{\pi g^4} [{\mathcal I}\bar{\mathcal I}] e^{-\frac{2m}{g^2}} \notag \\
&~=&
 -2m \left( \frac{g^2}{2m} \right)^{2(\epsilon-1)} \frac{\sin \epsilon \pi}{\pi} 
\Gamma \left( \epsilon \right)^2 e^{-\frac{2m}{g^2}} \times 
\left\{ 
\begin{array}{ll}
e^{\pi i \epsilon} ~&~ \mbox{for $\theta=-0$} \\
e^{-\pi i \epsilon} ~&~ \mbox{for $\theta=+0$}
\end{array} 
\right..
\label{eq:QMI}
\eeq
For $\epsilon \approx 1$, 
this non-perturbative correction to the ground state energy becomes
\beq
- \lim_{\beta \rightarrow \infty} \frac{1}{\beta} \frac{Z_1}{Z_0} &=&
-2m (\epsilon-1) e^{-\frac{2m}{g^2}}.
\eeq
This gives correct non-perturbative correction \eqref{eq:nearSUSY} 
in the near supersymmetric case $\epsilon \approx 1$. 
The result of the Gaussian approximation \eqref{eq:one-loop}
in the weak coupling limit $g \rightarrow 0$ 
agrees with Eq.\,\eqref{eq:QMI} 
if the gamma function is replaced by its asymptotic form
\beq
\Gamma( \epsilon ) \rightarrow \sqrt{\frac{2\pi}{\epsilon}} \epsilon^{\epsilon} e^{-\epsilon}.
\eeq 
Therefore, these two results agree in the large $\epsilon$ limit 
as expected from the discussion in Sec.\,\ref{sec:oneloop}.

\paragraph{On the saddle points of the original action and the effective bion potential \\}
As we have seen above, 
the saddle points of the effective bion potential 
$V_{\rm eff}$ are labeld by an integer $\sigma$. 
By comparing the values of the quasi moduli at each saddle points 
and those of the real bion \eqref{eq:rb_moduli} and 
the complex bion \eqref{eq:cb_moduli}, 
we find that the saddle points with $\sigma = 0$ and $\sigma = 1$ correspond 
to the weak coupling limit ($g \rightarrow 0$) of the real and the complex bions, respectively. 
Actually, the saddle point $\sigma = -1$ also corresponds to the complex bion. 
This is because the imaginary part of the kink positions
\beq
\tau_\pm = \tau_0 \pm \frac{\tau_r}{2}, 
\eeq
is defined modulo $2 \pi i /\omega \sim 2 \pi i / m$ 
and hence the following shift of the imaginary part does not change the physical configuration.
\beq
(m \tau_0 , m \tau_r) \sim (m \tau_0 - \pi i, m \tau_r + 2 \pi i).
\eeq
By using this shift, 
the value of $\tau_r$ at the saddle point with $\sigma=-1$ 
can be fixed to that for the complex bion. 
Nevertheless, the value of the action is different 
since the overall position $\tau_0$ has 
the imaginary part ${\rm Im} \, \tau_0 = -\pi i / m$, 
for which the integration path $C$ in \eqref{eq:Scb} has to be the line ${\rm Im} \tau = + \frac{1}{m}\frac{\pi i}{2}$.  

\section{Comments on bion contributions in sine-Gordon model}
\label{sec:SG}
As we have in the previous section, 
the sine-Gordon action \eqref{eq:SG_acton}, 
which can be obtained by restricting the $\C P^1$ action 
to the zero angular momentum sector, 
also has real and complex bion solutions \cite{Behtash:2015zha, Behtash:2015loa}. 
The crucial difference is that 
the bions in the sine-Gordon model do not have phase modulus. 
This is merely one manifestation of the fundamental difference of the topology of the target space: 
$S^1$ for the sine-Gordon model and 
${\mathbb C}P^1=S^2$ for the ${\mathbb C}P^1$ model. 
This fact particularly gives a marked difference 
when we consider quantum theory\footnote{
The measure of the wave function for the ${\mathbb C}P^1$ model 
is $\int d^2\varphi/(1+|\varphi|^2)^2
=\frac{1}{4}\int d\theta d\phi\sin\theta$, 
in contrast to $\int d\theta$ for sine-Gordon model.
}.
Consequently, the non-perturbative contributions to the ground state energy 
in the sine-Gordon model is different from that in the $\C P^1$ model. 
In the sine-Gordon model, the Gaussian approximation for the bion contributions 
which is valid in the limit $g \rightarrow 0$ with fixed 
$\lambda = m \epsilon g^2$, gives
\beq
- \lim_{\beta \rightarrow \infty} \frac{1}{\beta} \frac{Z_1}{Z_0} ~=~ 
 2\sqrt{\frac{8\omega^5}{\pi g^2 (\omega^2-m^2)}} \, (1+e^{\pm2 \pi i \epsilon}) \exp \left[ - \frac{2\omega}{g^2} - 2 \epsilon \log \frac{\omega+m}{\omega-m} \right],
\label{eq:SG_oneloop}
\eeq
while the complexified quasi moduli integral, 
which is valid in the limit $g \rightarrow 0$ with fixed $\epsilon$, 
gives 
\beq
- \lim_{\beta \rightarrow \infty} \frac{1}{\beta} \frac{Z_1}{Z_0} ~=~ 
\frac{m}{\pi} (1+e^{\pm2 \pi i \epsilon}) \, \Gamma(2\epsilon) \exp \left[ - \frac{2m}{g^2} + (2 \epsilon -1) \log \frac{g^2}{4m} \right],
\label{eq:SG_QMI}
\eeq
corresponding to $\theta = \arg \, g^2 =-, +$ for upper and lower sign respectively. 
These results do not agree with the corresponding 
non-perturbative corrections \eqref{eq:one-loop} and \eqref{eq:QMI} 
in the $\C P^1$ model. 
The mismatch of the ground state energies is due to 
the difference of the Hamiltonian $H_{\C P^1}$ in Eq.~(\ref{eq:H}) 
for the ${\mathbb C}P^1$ model and that obtained from the Lagrangian 
(\ref{eq:SG_acton}) for the sine-Gordon model
\beq
H_{\C P^1}^{l=0} &=& 
- g^2 \left( \p_\theta^2 + \frac{1}{\tan \theta} \p_\theta \right) + \frac{m^2}{4g^2} \sin^2 \theta - \epsilon m \cos \theta ~=~
H_{SG} - \frac{g^2}{\tan \theta} \p_\theta. 
\eeq
The nonperturbative corrections (\ref{eq:SG_oneloop}) 
and (\ref{eq:SG_QMI}) vanish in the limit $\epsilon =\frac{1}{2}$. 
This is in accord with the fact that $\epsilon =\frac{1}{2}$ is 
the supersymmetric limit of the sine-Gordon model.

As opposed to the $\C P^1$ model, 
the ambiguity in \eqref{eq:SG_oneloop} and \eqref{eq:SG_QMI} does not vanish 
in the near supersymmetric regime $\epsilon \approx \frac{1}{2}$. 
To compare it with the ambiguity in the perturbative part, 
let us consider the leading order correction 
to the ground state energy in the near supersymmetric limit:
\beq
E^{(1)} ~=~ E^{(1)}_{\rm pert} + E^{(1)}_{\rm bion},  
\eeq
where $E^{(1)}$ stands for the leading order coefficient 
in the small $\delta \epsilon \equiv \epsilon - \frac{1}{2}$ expansion of 
the ground state energy
\beq
E^{(1)} ~\equiv~ \lim_{\epsilon \rightarrow \frac{1}{2}} \p_\epsilon E.
\eeq
For $\epsilon = \frac{1}{2}$, the supersymmetric ground state wave function 
in the sine-Gordon quantum mechanics is given by
\beq
\Psi = \exp \left( \frac{m}{2g^2} \cos \theta \right).
\eeq
By using the standard perturbation theory with respect 
to small $\epsilon-\frac{1}{2} \equiv \delta\epsilon$, 
we obtain the correction to 
the ground state energy in the near supersymmetric case:
\beq
E ~\approx~ - m\, \delta\epsilon\, \frac{I_1 (m/g^2)}{I_0 (m/g^2)} 
~=~ - \delta\epsilon\, g^2 m \frac{\p}{\p m} \log I_0 (m/g^2), 
\eeq
where $I_1(m/g)$ and $I_0(m/g)$ are the modified Bessel function of the first kind.
The asymptotic expansion of the Bessel function
implies that the perturbative expansion gives 
the following asymptotic series for the correction to the ground state energy
\beq
E_{\rm pert}^{(1)} ~=~ - g^2 m \frac{\p}{\p m} \log e^{\frac{m}{g^2}} \sqrt{\frac{g^2}{2\pi m}} \left( 1 + \cdots + \frac{[(2n-1)!!]^2}{n!} \left( \frac{g^2}{8m} \right)^n + \cdots \right),
\eeq
The Borel resummation with negative $g^2$ and the analytic continuation to positive $g^2$ gives the ambiguity
\beq
E_{\rm pert}^{(1)} ~=~ - g^2 m \frac{\p}{\p m} \log \left[ I_0(m/g^2) \pm \frac{i}{\pi} K_0(m/g^2) \right],
\eeq
where $K_0(m/g^2)$ is the modified Bessel function of the second kind. 
Therefore, the non-perturbative part should have the following ambiguity
\beq
E_{\rm bion}^{(1)} ~=~ E^{(1)}-E_{\rm pert}^{(1)} 
~=~ g^2 m \frac{\p}{\p m} \log \left[1 \pm \frac{i}{\pi} \frac{K_0(m/g^2)}{I_0(m/g^2)} \right]. 
\eeq
The first leading order term is given by
\beq
E_{\rm bion}^{(1)} ~=~ 
g^2 m \frac{\p}{\p m} \log \left[1 \pm i e^{-\frac{2m}{g^2}} + \cdots \right] 
~=~ \mp 2i \, m \,e^{-\frac{2m}{g^2}} + \mathcal O \left( e^{-\frac{4m}{g^2}} \right),
\eeq
corresponding to $\theta \equiv {\rm arg} \, g^2 =-, +$ for upper and lower sign respectively. 
We can check that the complexified quasi moduli integral 
\eqref{eq:SG_QMI} is consistent with this near supersymmetric result.


\section{Summary and Discussion}
\label{sec:SD}
We have discussed the non-perturbative contributions from the complex saddle points 
in the ${\mathbb C}P^{N-1}$ and sine-Gordon quantum mechanics with the fermionic degrees of freedom.
We obtained non-perturbative contributions 
from the real and complex bion solutions
by using the Gaussian approximation, 
which is valid in the small coupling limit $g \rightarrow 0$ 
with fixed boson-fermion coupling constant $\lambda$ (large $\epsilon$ limit). 
For small $\epsilon$ including the supersymmetric ($\epsilon =1$) and the purely bosonic ($\epsilon=0$) cases, 
we investigated the integral along the Lefschetz thimbles associated with the saddle points 
to incorporate the contributions from the quasi zero modes, 
{\it i.e.}, the light normalizable mode around the real and complex bion solutions.
To evaluate the integrals along thimbles, 
we treated the bion configurations as well-separated instanton-antiinstanton configurations 
and calculate the quasi moduli integral 
with the complexified separation and phase
based on the Lefschetz thimble formalism.
The final results of the non-perturbative contributions from complexified saddle points
in the ${\mathbb C}P^{N-1}$ quantum mechanics are 
consistent with the known results for the supersymmetric case ($\epsilon = 1$) 
and the near supersymmetric case ($\epsilon \sim 1$).

Apart from the above main results, we have three more arguments:

(i) As discussed in Ref.~\cite{Behtash:2015loa}, 
we show that the result based on the Bogomolny--Zinn-Justin prescription, 
in which the sign of coupling $g^2$ is changed 
and is analytically continued back to the original sign in the final expression, 
is understood in terms of the complexification of the quasi moduli parameters 
both in the sine-Gordon quantum mechanics
and the ${\mathbb C}P^1$ quantum mechanics. 
In the complexified quasi moduli integral, 
we consider Lefschetz thimbles corresponding to decomposed cycles 
of the deformed integration contour. 
As we have discussed, these thimbles in the complexified 
quasi moduli integral are viewed as the approximated versions 
of the Lefschetz thimbles associated with 
the saddle points of the original complexified action.
To sum up, the quasi moduli integrals for instanton-antiinstanton configuration, 
which was originally performed based on the BZJ prescription, 
are nothing but the approximate versions of the Lefschetz thimble integrals 
associated with the real and complex bion saddle points in the complexified quantum mechanics. 
This is the reason why the imaginary part of the quasi moduli integral cancels 
that arising from the perturbative Borel resummation.

(ii) We elucidated the relation of ${\mathbb C}P^{N-1}$ field theory on ${\mathbb R}^{1}\times S^1$, 
and ${\mathbb C}P^{N-1}$ quantum mechanics and sine-Gordon quantum mechanics.
We showed that the ${\mathbb C}P^{N-1}$ sigma model 
on ${\mathbb R}^{1}\times S^{1}$ reduces to ${\mathbb C}P^{N-1}$ 
quantum mechanics by retaining modes with topological charge less than unity, 
but not to the sine-Gordon quantum mechanics.
We in particular established that bions in the two-dimensional ${\mathbb C}P^{N-1}$ field theory 
at small radius $L \ll 1$ is correctly described 
by means of ${\mathbb C}P^{N-1}$ quantum mechanics. 
These facts indicate that there is a smooth $L\to 0$ limit of the 
two-dimensional ${\mathbb C}P^{N-1}$ field theory and 
the resultant theory is correctly described by the ${\mathbb C}P^{N-1}$ quantum mechanics 
instead of the sine-Gordon quantum mechanics. 
After the renormalization procedure of the two-dimensional ${\mathbb C}P^{N-1}$ field theory, 
we expect that the most important part of the quantum dynamics 
should result in replacing the two-dimensional coupling 
by the running coupling $g_{\rm 2d}^2(1/L)$ with the renormalization scale $\mu=1/L$ 
\begin{equation}
\frac{1}{g_{\rm 1d}^2(1/L)}=\frac{L}{g_{\rm 2d}^2(1/L)}
\end{equation}
instead of Eq.(\ref{eq:1d2dCoupl}). 
More precise treatment including the quantum effects requires one-loop functional determinant 
with the appropriate background such as bions at the level of the field theory. 
This is one of the themes of our future works. 

(iii) We note that our real and complex bion solutions 
are the most general exact solutions of the equations 
of motion of the complexified ${\mathbb C}P^{1}$ quantum mechanics 
(\ref{eq:complexified_cp1}) (deformed by the fermion contributions) 
with the boundary condition (\ref{eq:BC}) on $R^1$ 
($-\infty< \tau < \infty$). 
This fact implies that no other exact solutions exist, including 
multiple bions on $R^1$ ($-\infty< \tau < \infty$). 
On the other hand, the resurgence theory requires 
that non-perturbative contributions with higher powers of 
non-perturbative exponential factors should exist, 
that are expected to come from multiple bion configurations. 
We anticipate that they should correspond to approximate 
solutions of the equations of motion which 
reduces to solutions asymptotically at large separations between bions in the complexified theory. 
This situation is quite similar to the single bion configuration 
in the ${\mathbb C}P^1$ quantum mechanics before the complexification. 
Hence our physical picture is the dilute gas of bions with 
short-range interactions. 
A similar picture has been advocated 
for the case of the sine-Gordon quantum mechanics\cite{Behtash:2015loa}. 
Alternatively, we anticipate to obtain exact solutions 
corresponding to multiple bions if we keep the compactification period
of the base space (inverse temperature) as $0 \le \tau < \beta$. 
We expect, however, the result should be 
the same as the above dilute gas picture in the large $\beta$ limit.

(iv) As for discussion,
there arises a question whether 
there exist 
complexified solutions playing similar roles 
in Yang-Mills or QCD. 
One possible connection is to consider 
$U(N)$ Yang-Mills theory coupled with Higgs fields 
in the fundamental representation.
In the Higgs phase, 
it allows a non-Abelian vortex whose 
low-energy dynamics is effectively described by 
the ${\mathbb C}P^{N-1}$ model localized around the vortex 
\cite{Hanany:2003hp,Auzzi:2003fs,Eto:2005yh}. 
When appropriate fermions are coupled in the original bulk theory, 
fermion quasi zero modes are also localized around the vortex and 
the ${\mathbb C}P^{N-1}$ model is coupled to the fermions.
If the bulk theory is supersymmetric, the vortex can be BPS 
and the supersymmetric 
${\mathbb C}P^{N-1}$ model is obtained as the vortex theory.
Therefore, our complexified solution should be able to be embedded into it 
either in non-supersymmetric or supersymmetric cases. 
Complexified bions should be able to be interpreted as those 
in (complexified) Yang-Mills theory in the bulk 
as instantons in ${\mathbb C}P^{N-1}$ model correspond 
to Yang-Mills instantons in the bulk \cite{Eto:2004rz}.
By taking a decoupling limit of the Higgs phase, 
we will be able to isolated complexified solutions in Yang-Mills theory.



\begin{acknowledgments}
We are grateful to Gerald Dunne and Mithat Unsal for the fruitful discussion
and thank the organizers for giving us a chance 
to discuss with Gerald in ``KEK Theory Workshop December 2015". 
This work is supported by
MEXT-Supported Program for the Strategic Research Foundation
at Private Universities ``Topological Science" (Grant No. S1511006).
This work is also supported in part 
by  the Japan Society for the 
Promotion of Science (JSPS) 
Grant-in-Aid for Scientific Research
(KAKENHI) Grant Numbers 
(16K17677 (T.\ M.), 16H03984 (M.\ N.) and 
25400241 (N.\ S.)).
The work of M.N. is also supported in part 
by a Grant-in-Aid for Scientific Research on Innovative Areas
``Topological Materials Science"
(KAKENHI Grant No. 15H05855) and 
``Nuclear Matter in neutron Stars investigated by experiments and
astronomical observations"
(KAKENHI Grant No. 15H00841) 
from the Ministry of Education, Culture, Sports, Science,
and Technology (MEXT) of Japan.

\end{acknowledgments}


\appendix

\section{Projection of fermionic degree of freedom}\label{appendix:fermion}
In this appendix, we derive the potential induced by projecting the fermionic degree of freedom. 
By redefining the fermionic degree of freedom as
\beq
\chi \equiv \frac{1}{g} \frac{1}{1+|\varphi|^2} \psi, \hs{10} \bar \chi \equiv \frac{1}{g} \frac{1}{1+|\varphi|^2} \bar \psi,
\eeq
the explicit form of the Lagrangian of the $\C P^1$ quantum mechanics \eqref{eq:L_QM} becomes
\beq
L = \frac{1}{g^2} \frac{|\dot \varphi|^2 - m^2 |\varphi|^2 }{(1+|\varphi|^2)^2} + i \bar \chi \left( \dot \chi - \frac{\bar \varphi \dot{\varphi}-\varphi \dot{\bar \varphi}}{1+|\varphi|^2} \chi \right) + \epsilon m \frac{1-|\varphi|^2}{1+|\varphi|^2} \bar \chi \chi.
\eeq
The corresponding classical Hamiltonian takes the form 
\beq
H = g^2 (1+|\varphi|^2)^2 \left[ |p_{\varphi}|^2 + \frac{i (\varphi p_{\varphi} - \bar \varphi \bar p_{\bar \varphi})}{1+|\varphi|^2} \chi \bar \chi \right] + \frac{m^2}{g^2} \frac{|\varphi|^2}{(1+|\varphi|^2)^2} - \epsilon m \frac{1-|\varphi|^2}{1+|\varphi|^2} \bar \chi \chi,
\eeq
where the conjugate momenta are given by
\beq
p_{\varphi} = \frac{\p L}{\p \dot \varphi} = \frac{\dot{\bar \varphi}}{(1+|\varphi|^2)^2}, \hs{10}
p_{\bar \varphi} = \frac{\p L}{\p \dot{\bar \varphi}} = \frac{\dot{\varphi}}{(1+|\varphi|^2)^2}, \hs{10}
-i \bar \chi = \frac{\p L}{\p \dot \chi}.
\eeq
The canonical commutation relations are
\beq
[\varphi, p_\varphi] = [ \bar \varphi, \bar p_{\bar \varphi} ] = i, \hs{10} \{\chi , \bar \chi\} = 1.
\eeq
Note that for $\epsilon =1$, the Hamiltonian commutes with the supercharges
\beq
Q &\equiv& g (1+|\varphi|^2) \left[ p_{\varphi} - \frac{1}{g^2} \frac{i m \bar \varphi}{(1+|\varphi|^2)^2} \right] \chi,  \\
\bar Q &\equiv& g (1+|\varphi|^2) \left[ p_{\bar \varphi} + \frac{1}{g^2} \frac{i m \varphi}{(1+|\varphi|^2)^2} - \frac{i}{2} \p_{\bar \varphi} \log G \right] \bar \chi, 
\eeq
which satisfy the supersymmetry algebra 
\beq
\{ Q , \bar Q \} = H + m q, 
\eeq
where $q$ is the conserved angular momentum
\beq
q \equiv i (\varphi p_{\varphi} - \bar \varphi p_{\bar \varphi}) + \chi \bar \chi .
\eeq

Since the fermion number operator commutes with the Hamiltonian
\beq
[H, \chi \bar \chi] =0 ,
\eeq
we can consider the projection onto the subspace of the Hilbert space with a fixed fermion number. 
By projecting onto the subspace with $\chi \bar \chi = 0$, i.e.
the sector annihilated by $\bar \chi$
\beq
\bar \chi | \Psi \rangle = 0,
\eeq
the Hamiltonian becomes 
\beq
H = g^2 (1+|\varphi|^2)^2 p_{\varphi} p_{\bar \varphi} + \frac{m^2}{g^2} \frac{|\varphi|^2}{(1+|\varphi|^2)^2} - \epsilon m \frac{1-|\varphi|^2}{1+|\varphi|^2}. 
\eeq
Therefore, the potential of the projected model is given by
\beq
V = \frac{m^2}{g^2} \frac{|\varphi|^2}{(1+|\varphi|^2)^2} - \epsilon m \frac{1-|\varphi|^2}{1+|\varphi|^2}. 
\eeq

\section{One-loop determinant}\label{appendix:determinant}
In this appendix, we derive the formula for the functional 
determinant \eqref{eq:determinant}. 
Let $\Delta$ be the following differential operator 
acting on $n$-component vectors 
\beq
\Delta = - \p_\tau^2 + V(\tau),
\eeq
where $V(\tau)$ is an $n$-by-$n$ matrix such that
\beq
\bar V = V, \hs{10} \lim_{\tau \rightarrow \pm \infty} V(\tau) 
~=~ M^2 ~~(\mbox{diagonal constant matrix}).  
\eeq
Here and in the following matrices with a bar (such as $\bar V$) 
denote their Hermitian conjugates .
We first consider the ratio of the determinants
\beq
D(\lambda) ~\equiv~ \frac{\det ( \lambda - \Delta \ )}
{ \det ( \lambda - \Delta_0) } ~=~ \prod_n 
\frac{\lambda - \lambda_n}{\lambda - \lambda_n^{0}},
\label{eq:defD}
\eeq
where $\Delta_0 = -\p_\tau^2 + M^2$ and 
$(\lambda_n,\, \lambda_n^0)$ are eigenvalues of 
$(\Delta, \Delta_0)$ respectively. 

From the formula $\log \det = \tr \log$, it follows that 
\beq
\frac{\p}{\p \lambda} \log D(\lambda) 
~=~ \tr \left[ \frac{1}{\lambda - \Delta} 
- \frac{1}{\lambda - \Delta_0} \right] 
~=~ - \int d\tau \, \tr\Big[ G(\tau, \tau) - G_0(\tau, \tau) \Big], 
\label{eq:logDelta}
\eeq
where $G$ and $G_0$ are the Green functions 
\beq
(\, \Delta \, - \, \lambda \,) G \ (\tau, \tau') 
&=& \mathbf 1_n \times \delta(\tau-\tau'), \label{eq:condG} \\ 
(\Delta_0 - \lambda) G_0(\tau,\tau') 
&=& \mathbf 1_n \times \delta(\tau-\tau'). 
\eeq
The Green function can be constructed 
from exponentially decreasing $n$-by-$n$ matrices 
$\psi_\lambda^{\pm}$ satisfying 
\beq
(\Delta - \lambda) \psi_\lambda^{\pm} = 0, \hs{10} 
\psi_\lambda^\pm ~\underset{\tau \rightarrow 
\pm \infty}{\longrightarrow}~ \exp( \mp \kappa \tau), 
\label{eq:asym_psi}
\eeq
where $\kappa$ is a diagonal matrix defined by
\beq
\kappa^2 = M^2 - \lambda ,
\eeq
which is positive definite for sufficiently small values of $\lambda$.
We can show that the Green function $G(\tau,\tau')$ is given by (see below)
\beq
G(\tau,\tau') ~=~ \psi_{\lambda}^+(\tau) \bar {\cal W}_\lambda^{-1} \bar \psi_{\lambda}^{-}(\tau') \theta(\tau-\tau') + \psi_{\lambda}^-(\tau) {\cal W}_\lambda^{-1} \bar \psi_{\lambda}^{+}(\tau') \theta(\tau'-\tau),
\label{eq:Green}
\eeq
where $\mathcal W_\lambda$ is the $n$-by-$n$ matrix 
defined as the Wronskian $W\big[ \psi_{\lambda}^+, \psi_{\lambda'}^- \big]$ with $\lambda' =\lambda$:
\beq
\mathcal W_\lambda = W\big[ \psi_{\lambda}^+, \psi_{\lambda}^- \big], \hs{10}
W\big[ \psi_{\lambda}^+, \psi_{\lambda'}^- \big] ~\equiv~ 
\bar \psi_{\lambda}^{+} \frac{\p}{\p \tau} \psi_{\lambda'}^{-} - \frac{\p}{\p \tau} \bar \psi_{\lambda}^{+} \, \psi_{\lambda'}^{-}. 
\eeq
Note that $\mathcal W_\lambda$ is independent of $\tau$
since $W\big[ \psi_{\lambda}^+, \psi_{\lambda'}^- \big]$ satisfies
\beq
\frac{\p}{\p \tau} W\big[ \psi_{\lambda}^+, \psi_{\lambda'}^- \big] &=& (\lambda-\lambda') \bar \psi_{\lambda}^{+} \psi_{\lambda'}^{-}. \phantom{\bigg[}
\label{eq:Wronskian}
\eeq
This relation also implies that the trace of the Green function at $\tau'=\tau$ can be rewritten as
\beq
\tr \, G(\tau,\tau) ~=~ \lim_{\lambda' \rightarrow \lambda} \frac{\p}{\p \tau} \tr \left( {\cal W}_\lambda^{-1} \frac{W \big[ \psi_{\lambda}^+, \psi_{\lambda'}^- \big]}{\lambda-\lambda'} \right).
\eeq
Similarly, the trace of the Green function of $\Delta_0$ is given by
\beq
\tr \, G_0(\tau,\tau) ~=~ \lim_{\lambda' \rightarrow \lambda} \frac{\p}{\p\tau} \tr \left( (2\kappa)^{-1} \frac{W \big[ e^{-\kappa \tau}, e^{\kappa' \tau} \big]}{\lambda-\lambda'} \right). 
\eeq
Using these expression, we can relate the determinant 
and the asymptotic forms of the Wronskians
\beq
\frac{\p}{\p \lambda} \log D(\lambda) &=& -\int_{-\infty}^{\infty} d\tau \, \tr \left[ G(\tau,\tau) - G_0(\tau,\tau) \right] \notag \\
&=& 
\lim_{T \rightarrow \infty} \lim_{\lambda' \rightarrow \lambda} 
\frac{1}{\lambda-\lambda'} \tr \bigg\{ (2\kappa)^{-1} W \big[ e^{-\kappa \tau}, e^{\kappa' \tau} \big] - {\cal W}_\lambda^{-1} W \big[ \psi_\lambda^+, \psi_{\lambda'}^- \big] \bigg\}_{\tau = -T}^{\tau=T}. 
\eeq
The asymptotic form of $W \big[ \psi_\lambda^+, \psi_{\lambda'}^- \big]$ can be 
obtained from those of $\psi^\pm_\lambda$. 
In addition to Eq.\,\eqref{eq:asym_psi}, 
let us assume the following asymptotic form in the opposite infinity
\beq
\psi_{\lambda}^+ &\longrightarrow& \exp( - \kappa \tau ) F^+(\lambda) + \exp( + \kappa \tau ) A^+(\lambda) ~~~~~\mbox{for $x \rightarrow - \infty$}, \\
\psi_{\lambda}^- &\longrightarrow& \exp( + \kappa \tau ) F^-(\lambda) + \exp( - \kappa \tau ) A^-(\lambda) ~~~~~\mbox{for $x \rightarrow +\infty$}, 
\eeq
where the coefficients $F^\pm(E)$ and $A^\pm(E)$ are $n$-by-$n$ matrices. 
Then the asymptotic forms of the Wronskian can be written as
\beq
W\big[ \psi_{\lambda}^+, \psi_{\lambda'}^-\big] ~=~ \left\{
\begin{array}{ll} 
(\kappa + \kappa') e^{-(\kappa - \kappa') \tau} F^-(\lambda') & \mbox{for $\tau \rightarrow + \infty$} \phantom{\Bigg[} \\
\overline{(\kappa + \kappa') e^{-(\kappa - \kappa') \tau} F^+(\lambda)} & \mbox{for $\tau \rightarrow - \infty$} \phantom{\Bigg[}
\end{array} \right..
\label{eq:aymptotic_W}
\eeq
In particular, for $\lambda'=\lambda$
\beq
{\cal W}_\lambda ~=~ 2 \kappa F^-(\lambda) ~=~ \overline{2 \kappa F^+(\lambda)}. 
\eeq
By using these asymptotic forms and
\beq
W \big[ e^{-\kappa \tau}, e^{\kappa' \tau} \big] = (\kappa + \kappa') e^{-(\kappa-\kappa')\tau}, 
\eeq
we find that
\beq
\frac{\p}{\p \lambda} \log D(\lambda) 
&=& \lim_{T \rightarrow \infty} \lim_{\lambda' \rightarrow \lambda} \tr \bigg[ (\kappa + \kappa') e^{-(\kappa - \kappa') T} 
\frac{F^{-}(\lambda) - F^{-}(\lambda')}{\lambda-\lambda'} \big\{ 2 \kappa F^{-}(\lambda) \big\}^{-1} \bigg] \notag \\
&=& \frac{\p}{\p \lambda} \log \det F^{-}(\lambda),
\eeq
where we have assumed that $\lambda < \lambda'$. 
This implies that $D(\lambda) \propto \det F^{-}(\lambda)$. 
Since $D(\infty) = 1$ (by definition) and $F^{-}(\infty) = \mathbf 1_n$ (free wave limit),
it follows that 
\beq
D(\lambda) = \det F^{-}(\lambda). 
\eeq

\paragraph{Removing zero modes \\}
Suppose that the operator $\Delta$ has $n$ zero modes. 
Let $\Xi_0$ be the basis ($n$-by-$n$ matrix) of the zero modes such that
\beq
\Xi_0 ~\rightarrow~ \exp( \mp M \tau) K^\pm ~~~~~ \mbox{as $\tau \rightarrow \pm \infty$},
\eeq
where $K^\pm$ are constant $n$-by-$n$ matrices. 
Note that the operator $\Delta$ can be rewritten as
\beq
\Delta \Xi_0 = (- \p_\tau^2 + V) \Xi_0 = 0 ~~~~~ \Longleftrightarrow ~~~~~ 
\Delta = - (\p_\tau + \p_\tau \Xi_0 \Xi_0^{-1}) (\p_\tau - \p_\tau \Xi_0 \Xi_0^{-1}). 
\eeq
The existence of $n$ normalizable zero modes $\Xi_0$ implies that
$\psi^\pm_\lambda$ become normalizable at $\lambda=0$ and hence 
\beq
F^\pm(\lambda = 0) = 0. 
\eeq
From the definition of $D(\lambda)$ in Eq.\,\eqref{eq:defD}, 
it follows that the zero mode can be removed by differentiating $D(\lambda)$: 
\beq
\frac{\det' \! \Delta \ }{\det \Delta_0} ~=~ \frac{(-1)^n}{n!} \frac{\p^n D}{\p \lambda^n} \bigg|_{\lambda=0} 
~=~ \det \left( - \frac{\p F^{-}}{\p \lambda \ } \right)_{\lambda=0},
\eeq
where $\det'$ stands for the determinant excluding the zero modes. 
By using the asymptotic behaviors of the Wronskian, 
the right hand side can be rewritten as
\beq
\frac{\p F^-}{\p \lambda \ } \bigg|_{\lambda=0} ~=~
\frac{1}{2M} \int_{-\infty}^{\infty} d\tau \frac{\p}{\p \tau} \frac{\p}{\p \lambda'} W\big[ \psi_0^+, \psi_{\lambda'}^- \big]_{\lambda'=0}. 
\label{eq:eq1}
\eeq
On the other hand, the Wronskian satisfies \eqref{eq:Wronskian}
and hence
\beq
\frac{\p}{\p \tau} \frac{\p}{\p \lambda'} W\big[ \psi_0^+, \psi_{\lambda'}^- \big]_{\lambda'=0} 
~=~ - \bar \psi_{\lambda=0}^+ \psi_{\lambda=0}^- 
=~ - (\bar K^+)^{-1} \bar \Xi_0 \Xi_0 (K^-)^{-1}. 
\label{eq:eq2}
\eeq
where we have used
\beq
\lim_{\lambda \rightarrow 0} \psi_\lambda^\pm = \Xi_0 (K^\pm)^{-1}.
\eeq
From Eqs.\,\eqref{eq:eq1} and \eqref{eq:eq2}, it follows that 
\beq
\frac{\det' \! \Delta \ }{\det \Delta_0} 
~=~ \det \left( - \frac{\p F^{-}}{\p \lambda \ } \right)_{\lambda=0} 
~=~ \frac{\det \mathcal G}{\det( 2M) \det \bar K^+ \det K^-}. 
\label{eq:theorem}
\eeq
where $\mathcal G$ is the matrix of the overlap integrals of zero modes
\beq
\mathcal G ~\equiv~ \int_{-\infty}^\infty d\tau \, \bar \Xi_0 \Xi_0.
\eeq

\paragraph{Green function \\}
Here we prove the formula for the Green function \eqref{eq:Green}. 
First note that $G(\tau, \tau')$ in Eq.\,\eqref{eq:Green} satisfies 
\beq
(\Delta - \lambda) G(\tau, \tau') = 0, \hs{10} \mbox{for $\tau \not = \tau'$}. 
\eeq
Therefore, the condition for the Green function \eqref{eq:condG}
is satisfied if
\beq
0 &=& \lim_{\tau' \rightarrow \tau+0} \ G(\tau, \tau') - \lim_{\tau' \rightarrow \tau-0} \ G(\tau, \tau'), \\
-\mathbf 1_n &=& \lim_{\tau' \rightarrow \tau+0} \p_\tau G(\tau, \tau') - \lim_{\tau' \rightarrow \tau-0} \p_\tau G(\tau, \tau'). 
\eeq

Define $\Omega$ by 
\beq
\Omega = - \p_\tau \psi_\lambda ^+ (\psi_\lambda ^{+})^{-1} 
+ \p_\tau \psi_\lambda ^{-} (\psi_\lambda ^{-})^{-1}.
\eeq
Then ${\cal W}_\lambda$ can be rewritten as
\beq
{\cal W}_\lambda ~=~ \bar \psi_\lambda ^+ \p_\tau \psi_\lambda ^{-} 
- \p_\tau \bar \psi_\lambda ^+ \psi_\lambda ^{-} 
&=& \bar \psi_\lambda ^{+} \left[ \Omega 
+ \p_\tau \psi_\lambda ^+ (\psi_\lambda ^{+})^{-1} \right] 
\psi_\lambda ^{-} - \p_\tau \bar \psi_\lambda ^+ \psi_\lambda ^{-} \notag \\
&=& \bar \psi_\lambda ^{+} \Omega \psi_\lambda ^{-} + W[\psi_\lambda ^+,\psi_\lambda ^+] (\psi_\lambda ^{+})^{-1} \psi_\lambda ^{-} \notag \\
&=& \bar \psi_\lambda ^{+} \Omega \psi_\lambda ^{-}.
\eeq
In the last equality, we have used
\beq
W[\psi_E^+,\psi_E^+] = 0, \hs{10} 
\left( \because \p_\tau W[\psi_E^+,\psi_E^+] 
= 0, ~~~ W[\psi_E^+,\psi_E^+] \underset{\tau \rightarrow + \infty}
{\longrightarrow} 0 \right).
\eeq
Then \eqref{eq:Green} can be rewritten as
\beq
G(\tau,\tau') 
= \Omega^{-1} \Big[( \bar \psi_\lambda^-(\tau))^{-1} 
\bar\psi_\lambda^{-}(\tau') \theta(\tau-\tau') 
+ (\bar \psi_\lambda^{+}(\tau))^{-1} \bar\psi_\lambda^{+}(\tau') \theta(\tau'-\tau) \Big] .
\eeq
By using this form of $G(\tau,\tau')$, we can easily show that
\beq
\lim_{\tau' \rightarrow \tau + 0} G(\tau,\tau') 
= \lim_{\tau' \rightarrow \tau - 0} G(\tau,\tau') = \Omega^{-1},
\eeq
and 
\beq
\lim_{\tau' \rightarrow \tau + 0} \p_\tau G(\tau,\tau') 
- \lim_{\tau' \rightarrow \tau - 0} \p_\tau G(\tau,\tau') 
=\Omega^{-1} \Big[(\bar \psi_\lambda^{+})^{-1} \p_\tau \bar\psi_\lambda^+ 
- (\bar\psi_\lambda^{-})^{-1}\p_\tau \bar\psi_\lambda^{-}  \Big]
= - \mathbf 1_n. 
\eeq
Therefore, $G(\tau,\tau')$ in Eq.\,\eqref{eq:Green} satisfies 
the condition for the Green function Eq.\,\eqref{eq:condG}.


\section{Measure for the moduli integral}
\label{appendix:measure}
In this appendix, we derive the measure for the moduli integral. 
Let us consider fluctuations $\xi^a$ around a classical 
background $\varphi_{\rm sol}^i(\eta^\alpha)$ as a function 
of the moduli parameters $\eta^\alpha$
\beq
\varphi^i = \varphi_{\rm sol}^i (\eta^\alpha) + e^i{}_a \xi^a,
\eeq
where $e^i{}_a$ are vielbein on the target space. 
The fluctuations are decomposed into zero modes $c^\alpha$ 
and massive modes $c^I$ 
\beq
\xi^a = \sum_{\alpha \in {\rm zero}\; {\rm modes}} c^\alpha \xi^a_{\alpha} 
+ \sum_{I \in {\rm massive}} c^I \xi^a_I. 
\eeq
The zero modes are related to the derivative with respect to the moduli parameters $\eta^\alpha$ as 
\beq
\xi_\alpha^a = e^a{}_i \frac{\p}{\p \eta^\alpha} \varphi_{\rm sol}^i. 
\eeq
The zero modes and massive modes are normalized as
\beq
\int d\tau \, \xi^a_\alpha \xi^a_\beta 
= {\mathcal G}_{\alpha \beta}, \hs{5}
\int d\tau \, \xi^a_I \xi^a_\alpha =0, \hs{5}
\int d\tau \, \xi^a_I \xi^a_J = \delta_{IJ},
\eeq
where $\mathcal G$ is the moduli space metric. 
The naive path integral measure is
\beq
\D \xi = \sqrt{\det \left( \frac{\mathcal G}{2\pi} \right)} 
\prod_{\alpha \in {\rm zero}\; {\rm modes}}  dc^\alpha \ \times \ 
\prod_{I \in {\rm massive}} \frac{\ dc^I}{\sqrt{2\pi}}. 
\eeq
To remove the zero mode integral, 
let us insert the following identity into the path integral:
\beq
1 &=& \int d^d \eta^\alpha \det 
\left( \int dx \, \xi_\alpha^a e^a{}_i \p_\beta \varphi^i \right) 
\delta^d \left( \int dx \, \xi_\alpha^a e^a{}_i \varphi^i \right) \\
&=& \int d^d \eta^\alpha \, \det \mathcal G \ 
\delta^d \left( \mathcal G_{\alpha \beta} c^\beta \right). 
\eeq
Integrating over $c^\alpha$, we obtain the measure for the moduli integral
\beq
\int \D \xi \, (\cdots) = \int d^d \eta^\alpha 
\sqrt{\det \left( \frac{\mathcal G}{2\pi} \right)} 
\prod_{I \in {\rm massive}} \frac{\ dc^I}{\sqrt{2\pi}} (\cdots) . 
\eeq


\end{document}